\documentclass[a4paper,fleqn,usenatbib]{mnras}

\usepackage[T1]{fontenc}
\usepackage{ae,aecompl}

\usepackage{graphicx}	
\usepackage{amsmath}	
\usepackage{amssymb}	

\renewcommand{\theenumi}{\arabic{enumi})}

\title[Focusing of Nonlinear Eccentric Waves]{Focusing of nonlinear eccentric waves in astrophysical discs}

\author[E. M. Lynch and G. I. Ogilvie]{
Elliot M. Lynch \thanks{E-mail: eml52@cam.ac.uk} and Gordon I. Ogilvie
\\
Department of Applied Mathematics and Theoretical Physics, University of Cambridge, Centre for Mathematical Sciences, \\ Wilberforce Road, Cambridge CB3 0WA, UK\\
}

\date{Accepted XXX. Received YYY; in original form ZZZ}

\pubyear{2019}

\begin{document}
\label{firstpage}
\pagerange{\pageref{firstpage}--\pageref{lastpage}}
\maketitle

\begin{abstract}
We develop a fully nonlinear approximation to the short-wavelength limit of eccentric waves in astrophysical discs, based on the averaged Lagrangian method of Whitham (1965). In this limit there is a separation of scales between the rapidly varying eccentric wave and the background disc. Despite having small eccentricities, such rapidly varying waves can be highly nonlinear, potentially approaching orbital intersection, and this can result in strong pressure gradients in the disc. We derive conditions for the steepening of nonlinearity and eccentricity as the waves propagate in a radially structured disc in this short-wavelength limit and show that the behaviour of the solution can be bounded by the behaviour of the WKB solution to the linearised equations. 
\end{abstract}




\section{Introduction}

Astrophysical discs can be eccentric in general, with the dominant motion of the fluid consisting of a sequence of confocal elliptical Keplerian orbits. These have many applications in astrophysics. They have been used to explain asymmetry and periodic variation of the emission lines of Be stars \citep{Ogilvie08} and white dwarfs \citep{Manser16,Cauley18,Miranda18}. They are also of potential interest in describing tidal disruption events where a star on a parabolic orbit around a supermassive black hole is torn apart by tidal forces. Hydrodynamic simulations of circumbinary discs \citep{MacFadyen08,Miranda17,Thun17,Thun18,Miranda18} consistently find that eccentricity is excited by the binary orbit and the tidally truncated inner disc edge can achieve significant eccentricities for a wide range of binary parameters. A closely related situation is the interaction of a disc with an embedded planet, which results in eccentricity being excited at the Lindblad resonances \citep{Papaloizou01,Goldreich03,Kley06,DAngelo06,Rice08,Duffell15,Teyssandier16,Teyssandier17,Rosotti17,Ragusa18}.

\citet{Ogilvie01} developed a theory of eccentric Keplerian discs based on a thin-disc expansion and orbital coordinates. Linearised theory for a 2D eccentric disc was developed by various authors \citep{Okazaki91,Savonije93,Okazaki97,Papaloizou02,Papaloizou06, Goodchild06}, who obtain different equations for the dynamics depending on how the vertical averaging is carried out (see  \citet{Ogilvie08} for a discussion). Linearised theory for 3D eccentric discs was derived by \citet{Ogilvie08} who found important differences in the dynamics of the eccentricity compared with 2D theory.

Unlike circular discs, the vertical structure of an eccentric disc cannot be decoupled from the horizontal fluid motion. Eccentric discs cannot maintain hydrostatic equilibrium owing to the variation of vertical gravity around an elliptical orbit. This drives a vertical ``breathing" mode where the disc scale height varies around the orbit; this variation results in pressure perturbations that affect the horizontal dynamics \citep{Ogilvie08}.

Nonlinear secular theory of eccentric discs has been developed by \citet{Ogilvie14}, \citet{Barker16} for an isothermal perfect gas using orbit averaged stress integrals.  

 WKB theory has been used by various authors to describe the behaviour of short-wavelength linear eccentric modes in the inner regions of discs. Examples of this approach are the work of \citet{Tremaine01} on galactic ``slow modes", \citet{Ferreira09} on eccentricity in the inner regions of black hole accretion discs and most recently \citet{Lee18} who looked at how the relative importance of pressure and self-gravity affects linear eccentric modes in discs. Of particular relevance to this paper, the theory of \citet{Ferreira09} predicts its own breakdown - with eccentric waves in the inner regions of black holes attaining very strong eccentricity gradients highlighting the need for a nonlinear theory to describe this regime.  \citet{Ferreira08} showed that the eccentricity gradients are important in determining the growth rates of inertial waves via a parametric type instability, which have been proposed as the origin of high frequency quasi-periodic oscillations (QPOs) \citep{Kato08,Ferreira08,Ferreira09,Dewberry18,Dewberry19}.

The extension of the short-wavelength limit into the nonlinear regime was first done by \citet{Lee99} who used the theory of \citet{Whitham65} to derive the dispersion relation for a single-armed spiral density wave in the tight-winding limit, incorporating pressure and self-gravity. They found for a range of surface density and sound speed profiles that a tightly wound trailing wave tends to become more nonlinear as it propagates inwards.

In \citet{Ogilvie19} we showed that the nonlinear equations of eccentric discs can be derived from a variational principle. This nonlinear secular theory describes the slow evolution of nested ellipses due to pressure forces in a thin disc. In the 3D theory the Hamiltonian is proportional to the orbit averaged internal energy of the disc. This formalism allows for a much simpler approach to the nonlinear dynamics of eccentric discs than that previously derived by \citet{Ogilvie01} and \citet{Ogilvie14} using stress integrals.

In this paper we apply the theory of \citet{Whitham65} to determine the behaviour of eccentric waves in the short-wavelength limit. Our work has much in common with that of \citet{Lee99} in the absence of self-gravity (neglected here). The main advantage of this work over \citet{Lee99} is that our short-wavelength limit is much less restrictive than their tight winding assumption, allowing for both untwisted discs (covered here) and twisted discs (to be addressed in future work). Our treatment of the pressure forces is more general than  \citet{Lee99}, in particular allowing for the inclusion of 3D effects. Our formalism should be readily extendable to include self-gravity, at which point the theory of \citet{Lee99} will be obtainable as a limit of this more general theory.

This paper is structured as follows. In Section \ref{2.5D overview} we discuss the geometry of eccentric discs and introduce the Hamiltonian formalism derived in \citet{Ogilvie19}. In Section \ref{low amp limit} we derive the equations for the short wavelength limit of an eccentric disc, and show that the equations take the form of a nonlinear oscillator on the short lengthscale. In Section \ref{short length dynamics} we solve for the dynamics of this nonlinear oscillator. We then apply the theory of \citet{Whitham65} to our short wavelength limit to obtain the modulation of the eccentric wave on the disc lengthscale (Section \ref{modulation}).We derive conditions on the behaviour of the modes in Section \ref{untwisted cond}. In Section \ref{results} we derive some representative solutions and in Section \ref{astro consequences} we discuss some general conclusions that can be drawn about eccentric waves in realistic astrophysical cases. Conclusions are given in Section \ref{conc} and mathematical derivations are given in the appendices.

\section{Nonlinear Theory of Eccentric Discs} \label{2.5D overview}

Let $(r,\phi)$ be polar coordinates in the disc plane. The polar equation for an elliptical Keplerian orbit of semimajor axis $a$, eccentricity $e$ and longitude of periapsis $\varpi$ is

\begin{equation}
r = \frac{a (1 - e^2)}{1 + e \cos f} \quad ,
\end{equation} 
where $f = \phi - \varpi$ is the true anomaly. A planar eccentric disc involves a continuous set of nested elliptical orbits. The shape of the disc can be described by considering $e$ and $\varpi$ to be functions of $a$. The derivatives of these functions are written as $e_a$ and $\varpi_a$, which can be thought of as the eccentricity gradient and the twist, respectively. The disc evolution is then described by the slow variation in time of the orbital elements $e$ and $\varpi$ due to secular forces such as pressure gradients in the disc and secular gravitational interactions.

In \citet{Ogilvie19} it was shown that the equation for the evolution of $e$ and $\varpi$ in a fluid disc can be obtained from a Hamiltonian formalism. Hamiltonian equations for an eccentric disc in terms of $e$ and $\varpi$ are

\begin{equation}
M_a \dot{e} = \frac{\sqrt{1 - e^2}}{n a^2 e} \frac{\delta H}{\delta \varpi} \quad ,
\label{imag ham}
\end{equation}

\begin{equation}
M_a \dot{\varpi} = -\frac{\sqrt{1 - e^2}}{n a^2 e} \frac{\delta H}{\delta e} \quad,
\label{re ham}
\end{equation}
where $M_a = dM/da$ is the one-dimensional mass density with respect to $a$, $n = (G M_{\star}/a^3)^{1/2}$ is the mean motion, $M_{\star}$ is the central mass and $H$ is the Hamiltonian. A dot denotes a partial derivative with respect to time. Writing the Hamiltonian as an integral of the Hamiltonian density $H_a$:

\begin{equation}
H = \int H_{a} \, da \quad ,
\end{equation}
the functional derivatives are given by

\begin{equation}
\frac{\delta H}{\delta \varpi} = \frac{\partial H_a}{\partial \varpi} - \frac{\partial}{\partial a} \frac{\partial H_a}{\partial \varpi_a} \quad ,
\end{equation}

\begin{equation}
\frac{\delta H}{\delta e} = \frac{\partial H_a}{\partial e} - \frac{\partial}{\partial a} \frac{\partial H_a}{\partial e_a} \quad .
\end{equation}
Equations (\ref{imag ham}) and (\ref{re ham}) form a system of nonlinear PDEs which are 2nd order in space that describe a type of nonlinear dispersive wave. 

The Hamiltonian of the 3D and 2D theories differ and this has important dynamical consequences. In a disc where the only non-Keplerian terms are those from pressure gradients, both the 2D and 3D Hamiltonians are proportional to the disc internal energy. From \citet{Ogilvie19} the Hamiltonian density of the 2D theory is

\begin{equation}
H_{a}^{\rm (2D)} = M_a \langle \varepsilon \rangle \quad ,
\end{equation}
where $\varepsilon$ is the specific internal energy of the disc, and (in this section only) the angle brackets denote an orbit average. For a 3D disc the Hamiltonian is modified because of the presence of the disc breathing mode or scale height variation around the orbit. The Hamiltonian density is then

\begin{equation}
H_{a}^{(3\mathrm{D})} = \frac{1}{2} (\gamma + 1) M_a \langle \bar{\varepsilon} \rangle \quad ,
\end{equation}
where $\bar{\varepsilon}$ is the mass-weighted vertically averaged specific internal energy. Both the 2D and 3D Hamiltonian densities can be written as $H_a = H^{\circ}_a F$ where $H^{\circ}_a = 2 \pi a P^{\circ}$ is the Hamiltonian of a circular disc, with $P^{\circ}$ the radial profile of the vertically integrated pressure in the limit of a circular disc, and $F = F(e, a e_a, a \varpi_a)$ is a dimensionless function that depends on the disc geometry \citep{Ogilvie19}.

Additional precessional forces can be included by adding a term to the Hamiltonian that depends on $a$ and $e^2$ only: $H^{f}_{a} = H^{f}_a (a,e^2)$, so that the Hamiltonian density is given by $H_a = H_{a}^{\rm (2D)}  + H^{f}_{a}$ for a 2D disc and $H_a = H_{a}^{\rm (3D)}  + H^{f}_{a}$ for a 3D disc.
 
There are two sources of nonlinearity present in the Hamiltonian density for both the 2D and 3D theories. One is associated with the eccentricity $e$; the other is associated with the gradients of the solution and typically determines how oscillatory (or twisted) the wave is, with more oscillatory (or twisted) waves being more nonlinear. To characterise this form of nonlinearity we introduce a nonlinearity parameter $q$ given by

\begin{equation}
q^2 = \frac{(a e_a)^2 + (1 - e^2) (a e \varpi_a)^2}{[1 - e (e + a e_a)]^2} \quad ,
\end{equation}
which matches the definition in \citet{Ogilvie19} (their Appendix C). We require $|q| < 1$ to avoid an orbital intersection \citep{Ogilvie19}. The relative contribution of the eccentricity gradient and twist to $q$ is determined by an angle $\alpha$ with 

\begin{equation}
\frac{a e_a}{1 - e(e + a e_a)} = q \cos \alpha \quad. 
\end{equation}
In these variables the geometric part of the Hamiltonian can be written as $F = F(e,q,\alpha)$.

The simplest solutions to Equations (\ref{imag ham}) and (\ref{re ham}) are obtained by requiring that the complex eccentricity $E = e \exp{i \varpi}$ depends on time only through a phase factor so that $E \propto \exp {i \omega t}$. This corresponds to a solution that is steady in some rotating frame, i.e. a uniformly precessing eccentric disc. When $E$ takes this form, the twisted solutions are eccentric travelling waves whilst the untwisted solutions are the standing waves or modal solutions of the disc. The equations for these eccentric waves are

\begin{equation}
0 = \frac{\delta H}{\delta \varpi} \quad ,
\label{imag ham mode}
\end{equation}

\begin{equation}
M_a \omega = -\frac{\sqrt{1 - e^2}}{n a^2 e} \frac{\delta H}{\delta e} \quad ,
\label{re ham mode}
\end{equation}
where $\omega = \dot{\varpi}$ is the global precession rate of the disc.

The untwisted eccentric modes can be thought of as a superposition of two twisted eccentric waves which reflect off the disc boundaries (such as the free edge considured in \citet{Ogilvie19}) or are potentially confined away from the disc boundary in a wave cavity \citep{Lee19}. In the presence of an excitation mechanism (such as an eccentric Lindblad resonance with an orbital companion, as occurs in superhump binaries, for example) we would expect a reflective disc to develop an untwisted eccentric mode. However, if the  ingoing and outgoing travelling waves have different amplitudes, for instance if the boundaries are not perfectly reflecting or if the dissipation is strong enough, then the disc will develop a twist. 

Equations (\ref{imag ham mode}) and (\ref{re ham mode}) can be derived from the variation of the Lagrangian

\begin{equation}
L =  \omega J - H \quad ,
\label{general mode lagrangian}
\end{equation}
where (in this section only) $J$ is the total angular momentum in the disc and is given by

\begin{equation}
 J = \int M_{a} n a^2 \sqrt{1 - e^2} \, d a \quad .
\end{equation}

As noted in \citet{Ogilvie19}, in the untwisted case the variation of Equation (\ref{general mode lagrangian}) corresponds to seeking stationary values of $H$ subject to the constraint $J = \mathrm{constant}$, with the precession frequency appearing as a Lagrange multiplier. 

The 3D Hamiltonian in \citet{Ogilvie19} can be obtained as the long-timescale dynamics of an averaged Lagrangian (in the sense of \citet{Whitham65}) where the average is taken over the orbital timescale.

The precessional forces coming from $H^{f}_a$ can be expressed in terms of the precession rate of a test particle

\begin{equation}
M_{a} \omega_{f} (a,e) := -\frac{\sqrt{1 - e^2}}{n a^2 e} \frac{\partial H^{f}_a}{\partial e} \quad ,
\end{equation}
 which can be rearranged for $H^{f}_a$,

\begin{equation}
H^{f}_a = -M_{a} n a^2 \int \frac{e \omega_{f} (a,e) }{\sqrt{1 - e^2}} \, d e 
\end{equation}
and the Lagrangian becomes:

\begin{equation}
L = \int \left[ M_{a} n a^2 \left( \omega \sqrt{1 - e^2} + \int  \frac{e \omega_{f} (a,e) }{\sqrt{1 - e^2}} \, d e \right)  - H^{\circ}_a F \right]\, da \quad .
\label{lagrangian for expansion}
\end{equation}

\section{Short-Wavelength Nonlinear Eccentric Waves Using an Average Lagrangian} \label{low amp limit}

We now consider  the situation where the precessional (or test particle) forces are strong relative to the pressure forces. Defining

\begin{equation}
N^2 = \frac{M_a n a^2 ( \omega_{f} (a,0) - \omega)}{H^{\circ}_{a}} \quad ,
\end{equation}
which can be understood as the ratio of these precessional effects (as captured by $\omega_{f} (a,0) - \omega$) to pressure forces, this situation corresponds to the limit $N^2 \gg 1$. This limit can come about either because of a prograde externally induced precession ($\omega_f(a,0)>0$, e.g. due to general relativity or an orbital companion), or from a retrograde free precession of the disc ($\omega<0$). \footnote{The correspondence of $N^2 \gg 1$ to the short wavelength limit can be seen in linear theory. For example, the linear, local dispersion relation for an isothermal disc is $(k_r r)^2 = 2 N^2$ where $k_r$ is the radial wavenumber. As we shall show, stongly nonlinear effects can break this correspondence.}

The method of \citet{Whitham65} is applicable to nonlinear waves propagating on a background varying on a much longer lengthscale than the wavelength of the wave (it is also applicable to temporal variations but this is not needed here). A separation of scales occurs, as over one cycle, the wave sees a constant background, with the variations in the background medium only having an effect over many wave cycles. The wave  amplitude is modulated on the longer lengthscale by this variation in the background medium.

In order to use the theory of \citet{Whitham65} we introduce a rapidly varying phase variable $\varphi = \varphi(a)$ which satisfies $a \varphi_a = N(a) k(a)$ with $k(a) = O(1)$ a rescaled dimensionless wavenumber. We can then introduce a rescaled variable $\tilde{e} = O(1)$ such that

\begin{equation}
e = \frac{\tilde{e} (\varphi,a)}{N(a) k(a)} \quad .
\end{equation}
For an untwisted disc the nonlinearity becomes

\begin{equation}
q = a \frac{d e}{d a} = \tilde{e}_{\varphi} (\varphi,a) + a \frac{\partial}{\partial a} \left( \frac{\tilde{e} (\varphi,a)}{N(a) k(a)} \right) \approx \tilde{e}_{\varphi} (\varphi,a) \quad .
\end{equation}
which is of order unity.

Substituting this into the Lagrangian (Equation \ref{lagrangian for expansion}) and keeping only terms of the lowest order in $N^{-1}$, the Lagrangian becomes

\begin{equation}
L \approx  \int H^{\circ}_a \hat{\mathcal{L}} \, d a= \int H^{\circ}_a  \left(\frac{1}{2} \frac{\tilde{e}^2}{k^2}  - F(\tilde{e}_{\varphi}) \right) \, da \quad ,
\label{lagrangian in limit}
\end{equation}
where we have neglected a term ($M_{a} n a^2 \omega$) which is a function only of $a$ as this has no effect on the dynamics. We have separated the Lagrangian into a dimensionless part $\hat{\mathcal{L}}$ which controls the small scale dynamics and a dimensional part $H^{\circ}_{a}$ which is important in determining how the solution varies on the length scale of the disc. 

We show in Appendix \ref{geo F deriv} why at lowest order $F(e,q,\alpha) \approx F(q)$. The form of $F(q)$ depends on the ratio of specific heats and whether the disc considered includes the effects of the breathing mode. Example forms for $F(q)$ are

\begin{align}
F^{(\rm lin)} (q) &= \frac{\gamma + 1}{2 (\gamma - 1)} + \frac{(2 \gamma - 1)}{4 \gamma} q^2 \quad ,\\
F^{(\rm 2D)} (q) &= \frac{1}{(\gamma - 1)} \frac{1}{2 \pi} \int^{2 \pi}_{0} (1 - q \cos E)^{-(\gamma - 1)} d E \quad , \\
F^{(\rm 3D)} (q) &=\frac{(\gamma + 1)}{2 (\gamma - 1)}  \frac{1}{2 \pi} \int^{2 \pi}_{0} [ h(q,\cos E) (1 - q \cos E) ]^{-(\gamma - 1)} d E \quad , \\
F^{(\rm iso)} (q) &= \ln \left[ 2 q^{-2} (1 - \sqrt{1 - q^2})\right] \quad , 
\end{align}
where $F^{(\rm lin)}$ is for a 3D adiabatic disc in linear theory, $F^{(\rm 2D)}$ and $F^{(\rm 3D)}$ are for the 2D and 3D adiabatic discs and $F^{(\rm iso)}$ is for an isothermal disc. Here $h(q,\cos E)$ is the dimensionless scale height (equal to the scale height divided by the orbit averaged scale height); this varies around the orbit owing to the presence of the disc breathing mode which is forced by the variation of the vertical gravity around the orbit (\citet{Ogilvie01},\citet{Ogilvie08},\citet{Ogilvie14}).

At lowest order there is no difference between the 3D and 2D isothermal discs, as the breathing mode is independent of $q$ owing to the vertical specific enthalpy gradient being independent of horizontal compression. The 3D isothermal disc will have an $O(1)$ contribution to $N^2$.

\begin{figure}
\includegraphics[width=\linewidth]{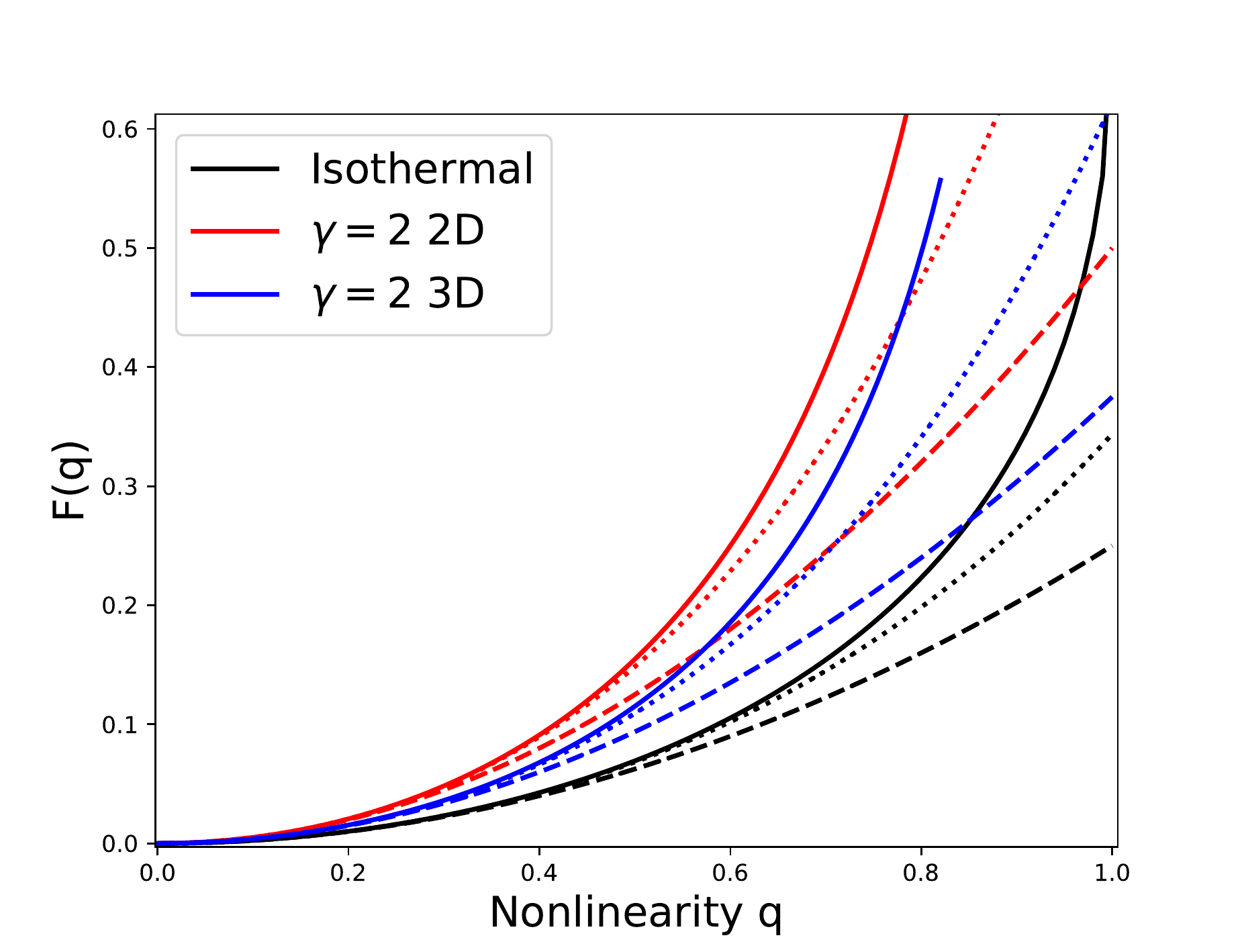}
\caption{Geometric part of the Hamiltonian in the short-wavelength limit. In each case the unimportant constant term has been subtracted off. Most forms of $F$ of interest are bracketed by the behaviour of these functions. In particular we expect all should have the property of being monotonically increasing even functions which diverge at $q=1$. The dashed lines are series expansion in $q$ including terms up to $O(q^2)$ (which corresponds to linear theory). The dotted lines includes $O(q^4)$ terms which are the lowest order nonlinear terms.}
\end{figure}

\begin{figure}
\includegraphics[width=\linewidth]{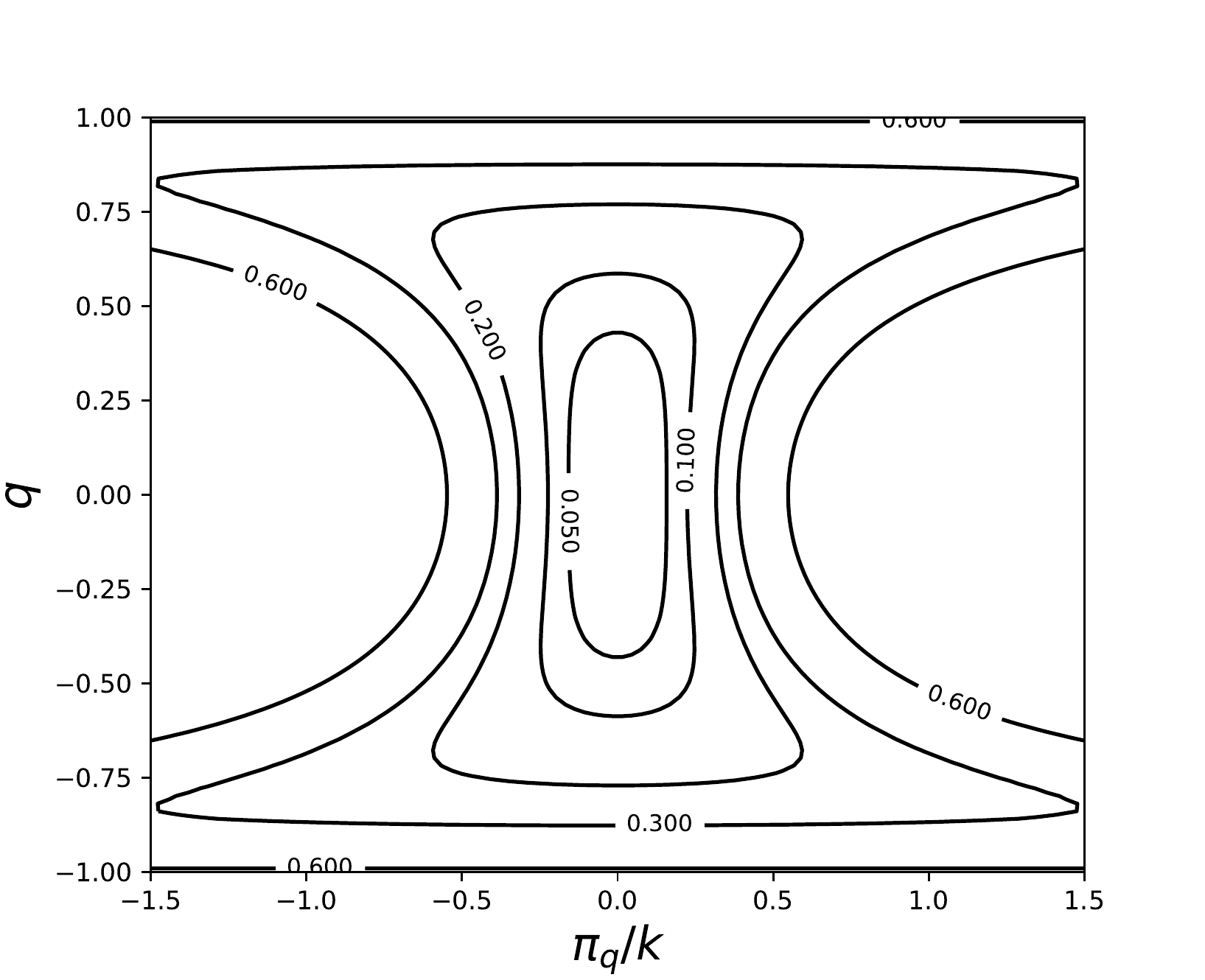}
\caption{Phase space of the small scale nonlinear oscillator for $F = F^{(\rm iso)}$ showing the Hamiltonian in terms of $q$ and $\pi_q/k$. }
\label{untwisted hamiltonian contours}
\end{figure}

It is useful to reformulate the Lagrangian viewing $q$ as the coordinate and obtain a Hamiltonian in terms of $q$ and its conjugate momentum $\pi_q$ as we expect this to be quadratic in $\pi_q$. To do so we vary $\hat{\mathcal{L}}$ with respect to $\tilde{e}$ to obtain the equation of motion

\begin{equation}
\tilde{e} = - k^2 F^{\prime \prime} (\tilde{e}_{\varphi}) \tilde{e}_{\varphi \varphi} 
\label{e eom}
\end{equation}
and thus rewrite the nondimensional Lagrangian as

\begin{equation}
\hat{\mathcal{L}} = \frac{k^2}{2} (F^{\prime \prime} (q) )^2 q_{\varphi}^2 - F(q)
\label{q lagrangian}
\end{equation}
where we have made use of the fact that $q = \tilde{e}_{\varphi}$. From this we obtain $\pi_q$:

\begin{equation}
\pi_q = \frac{\partial \hat{\mathcal{L}}}{\partial q_{\varphi}} = k^2 (F^{\prime \prime} (q))^2 q_{\varphi} \quad ,
\end{equation}
and thus obtain the Hamiltonian

\begin{equation}
H = \frac{\pi_q^2}{2 k^2 (F^{\prime \prime} (q) )^2} + F(q) \quad .
\label{q Hamiltonian}
\end{equation}

As we are treating $\varphi$ as a phase variable governing the short length scale behaviour we require $\tilde{e}$ to be periodic in $\varphi$ with period $2 \pi$. The dynamics on the short length scale can be entirely understood from the contours of constant Hamiltonian in the $(q, \pi_q/k)$ plane (Figure \ref{untwisted hamiltonian contours}) where they circulate around the fixed point at the origin. 

Throughout this paper we will need to make use of certain functionals of $F$ which are independent of the details of the disc model; these functionals and some of their properties are detailed in Appendix \ref{integral proofs}. For the long lengthscale dynamics of an untwisted disc the most important of these functionals are:

\begin{equation}
I (q_{+}) :=  \int_{0}^{q_{+}} F^{\prime \prime} (q) [F(q_{+}) - F(q)]^{-1/2} d q \quad ,
\label{complete untwisted I}
\end{equation}

\begin{equation}
J (q_{+}) := \int_{0}^{q_{+}} F^{\prime \prime} (q) [F(q_{+}) - F(q)]^{1/2} d q \quad .
\label{complete untwisted J}
\end{equation}

\section{Short-lengthscale dynamics} \label{short length dynamics}

With the preliminaries out of the way we now turn to solving the short-lengthscale dynamics using the Hamiltonian given by equation (\ref{q Hamiltonian}). Following \citet{Whitham65} we set $H = A$, where $A$ can be viewed as the amplitude of the nonlinear wave.  If $F$ is an even increasing function of $|q|$ then the maximum value of $q$ is attained when $\pi_q = 0$. Denoting this maximum value $q_{+}$ such that $|q| \le q_{+}$ then we have $A = F(q_{+})$ so that 

\begin{equation}
 F(q_{+}) =  \frac{\pi_q^2}{2 k^2 (F^{\prime \prime} (q) )^2} + F(q) \quad .
\end{equation}
This can be rearranged to give

\begin{equation}
q_{\varphi} = \pm \frac{\sqrt{2}}{ k} \frac{\left[F(q_{+}) - F(q) \right]^{1/2}}{F^{\prime \prime} (q)} \quad ,
\end{equation}
and directly integrated to obtain an implicit solution for the short-lengthscale dynamics 

\begin{equation}
\pm \frac{\varphi}{\sqrt{2} k}  =  \int_{-q_{+}}^{q} \frac{F^{\prime \prime} (q)}{2 \left[F(q_{+}) - F(q) \right]^{1/2}} \, d q=I \left(q, q_{+} \right) \quad ,
\end{equation}
which can be written in terms of the incomplete form of $I$ (Equation \ref{incomplete untwisted I}) given in Appendix \ref{integral proofs}. Here the sign denotes the increasing and decreasing branches of the solution and is necessary as $\varphi(q)$ is multivalued over one period.

$k$ can be determined by imposing $2 \pi$ periodicity on $q(\varphi)$. With this condition we obtain the nonlinear dispersion relation for the rescaled wavenumber

\begin{equation}
k = \frac{\pi}{\sqrt{2} I \left(q_{+}, q_{+} \right)}  = \frac{\pi}{\sqrt{2} I \left( q_{+} \right)}  \quad ,
\end{equation}
allowing us to write

\begin{equation}
\varphi = \pm \pi \frac{ I \left(q, q_{+} \right)}{ I \left(q_{+} \right)} \quad .
\end{equation}

\begin{figure}
\includegraphics[width=\linewidth]{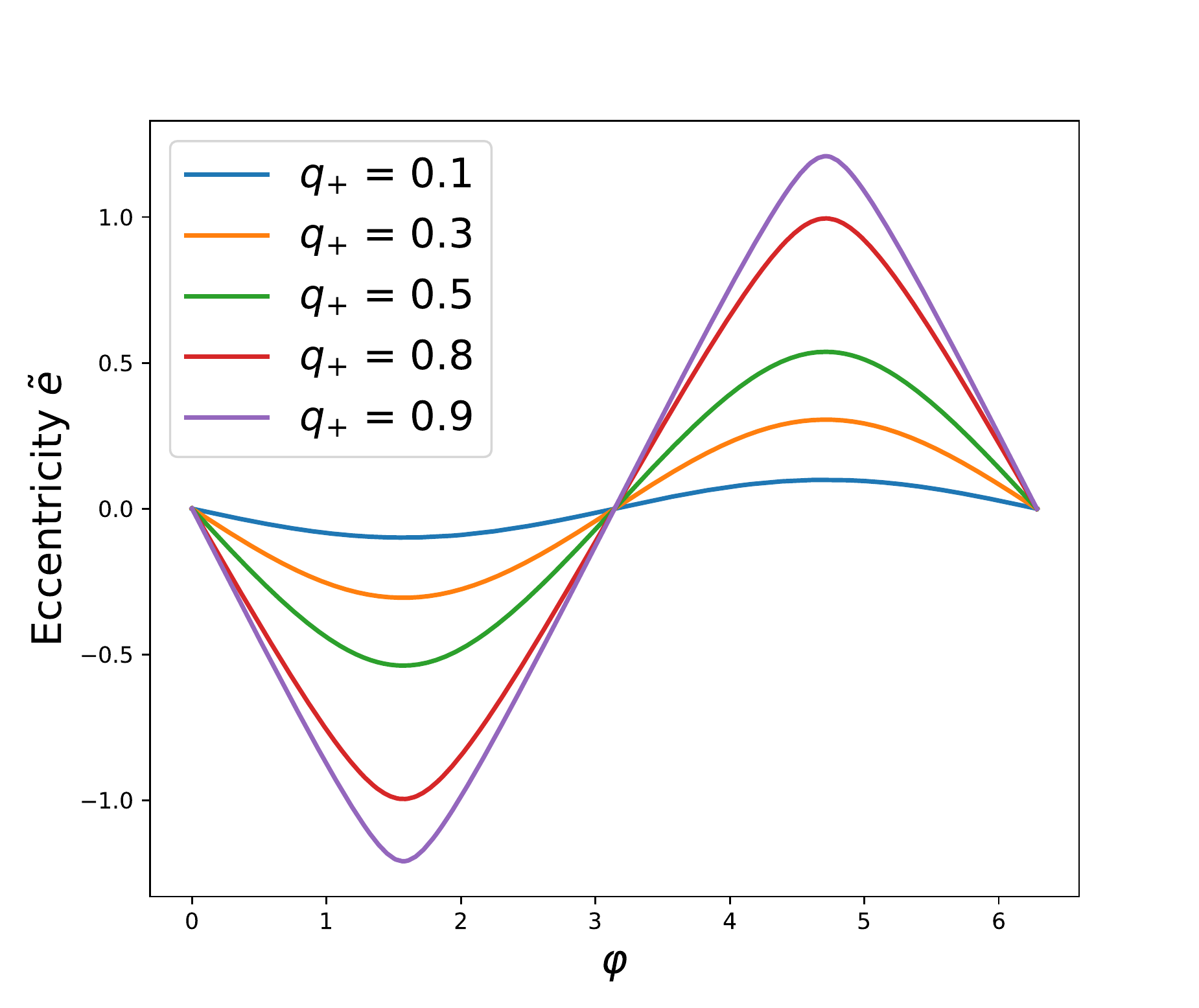}
\includegraphics[width=\linewidth]{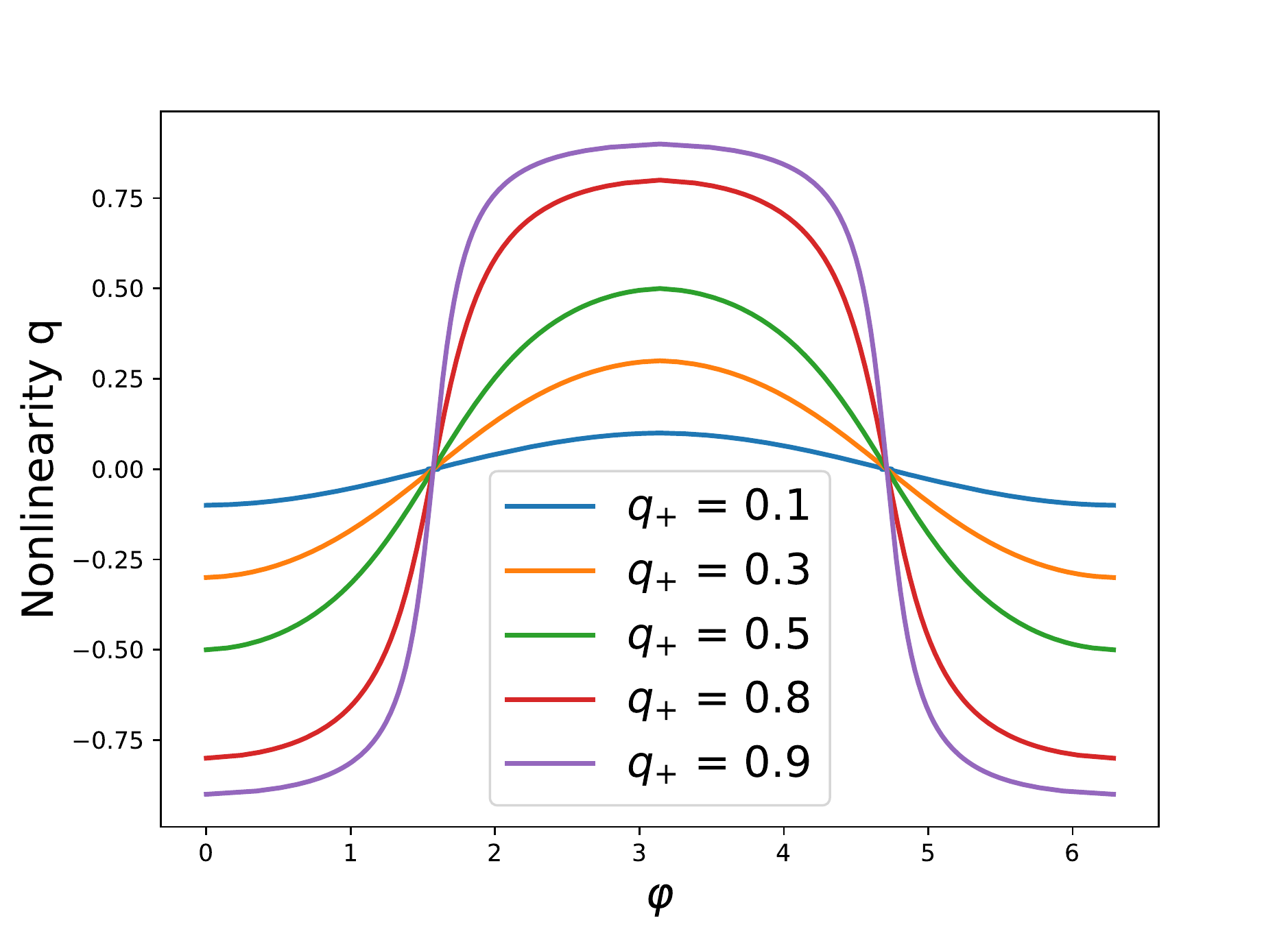}
\caption{Variation of the eccentric mode on the short lengthscale ($\varphi$) for different maximum nonlinearities ($q_{+}$) in an isothermal disc. Top (rescaled) eccentricity, bottom nonlinearity $q$. The eccentricity tends to a triangle wave and the nonlinearity to a square wave in the nonlinear limit $q_{+} \rightarrow 1$.}
\label{short length fig}
\end{figure}

Figure \ref{short length fig} shows the variation of the mode on the short lengthscale for an isothermal disc with different values of $q_{+}$. In the linear limit the eccentricity and nonlinearity are sine and cosine waves; as the nonlinear limit is approached they approach a limiting slope resulting in $e$ taking the form of a triangle wave, while $q$ tends to a square wave. This is similar to the nonlinear global modes calculated by \citet{Barker16} and \citet{Ogilvie19} using the more general theory. The advantage here is that it is easy to obtain the locations of the zeros and peaks of these modes which is not straightforward to obtain in general \citep{Ogilvie19}.

\section{Modulation of the eccentric wave on the disc length scale} \label{modulation}

\subsection{Averaged Lagrangian theory of  \citet{Whitham65}} 

In the method of \citet{Whitham65} the dynamics of the long lengthscale can be obtained from the variation of a Lagrangian obtained by averaging the full Lagrangian over one wave cycle. The long-lengthscale dynamics are obtained in the form of a set of conserved fluxes which the wave carries through the background medium. Conservation of these fluxes as the background varies results in variations in the wave amplitude.

The method of \citet{Whitham65} can be viewed as a method of obtaining the wave action and wave action fluxes for a nonlinear wave on a slowly varying background. Alternatively the conserved fluxes can be thought of as the Noether currents generated by the (approximate) symmetry of the Lagrangian under a change of phase.

It is easier to apply the method of \citet{Whitham65} to the Lagrangian with $q$ as a coordinate (equation \ref{q lagrangian}). This is because when $q$ is regarded as a coordinate the nonlinearity occurs in the potential energy term, whereas when $\tilde{e}$ is the coordinate this nonlinearity is in the kinetic energy term instead, which is problematic when inverting the canonical momenta.

The averaged Lagrangian is given by 

\begin{equation}
\langle L \rangle := \frac{1}{2 \pi} \int_{0}^{2 \pi} L \, d \varphi = \int H^{\circ}_a \langle \hat{\mathcal{L}} \rangle \, d a \quad, 
\end{equation}
where here the angle brackets denote an average over the short length scale. The (dimensionless) averaged Lagrangian $\langle \hat{\mathcal{L}} \rangle$ depends on the wave amplitude $A$ and the rescaled wavenumber $k$. Varying with respect to the wave amplitude $A$, we obtain the dispersion relation

\begin{equation}
\frac{\delta \langle L \rangle}{\delta A} = H^{\circ}_a \frac{\partial \langle \hat{\mathcal{L}} \rangle}{\partial A} = 0 \quad ,
\label{amplitude var}
\end{equation}
while varying with respect to $\varphi$, the short length scale variable, gives us a flux conservation equation

\begin{equation}
\frac{\delta \langle L \rangle}{\delta \varphi} = -\frac{\partial}{\partial a} \left( \frac{a H^{\circ}_a}{N} \frac{\partial \langle \hat{\mathcal{L}} \rangle}{\partial k} \right) = 0 \quad ,
\end{equation}
which corresponds to a conservation law

\begin{equation}
\frac{a H^{\circ}_a}{N} \mathcal{J} [A] = \mathrm{const}
\end{equation}
where $\mathcal{J}  := \frac{\partial \langle \hat{\mathcal{L}} \rangle }{\partial k}$. This corresponds to an algebraic equation which gives the variation of the amplitude $A$ throughout the disc. 

\subsection{Flux conservation and nonlinear dispersion relation for eccentric waves}

As per \citet{Whitham65} by setting $H = A$ we obtain an equation for $\pi_q$ in terms of $q$ and $A$,
 
\begin{equation}
\pi_q = \pm \sqrt{2} k F^{\prime \prime} (q) [A - F(q)]^{1/2},
\end{equation}
where this is left in terms of $A$ (rather than $F(q_{+})$) so that we can obtain the dispersion relation from  equation (\ref{amplitude var}). From this we obtain an expression for the (nondimensional) averaged Lagrangian,

\begin{align}
\begin{split}
\langle \hat{\mathcal{L}} \rangle &= \frac{1}{2 \pi} \int \pi_{q} q_{\varphi} \, d \varphi - H \\
&= \frac{1}{2 \pi} \sqrt{2} k \oint F^{\prime \prime} (q) [A - F(q)]^{1/2} d q - A \quad ,
\end{split}
\end{align}
where $\oint$ denotes an integral over one cycle in $q$. Varying with respect to the amplitude $A$ we obtain 

\begin{equation}
\langle \mathcal{L} \rangle_A = H^{\circ}_a \left( \frac{1}{2 \sqrt{2} \pi} k \oint F^{\prime \prime} (q) [A - F(q)]^{-1/2} d q - 1 \right) = 0,
\end{equation}
from which we obtain (after rescaling) the dispersion relation

\begin{equation}
a \varphi_a = \frac{2 \sqrt{2} \pi N}{\oint F^{\prime \prime} (q) [A - F(q)]^{-1/2} d q } \quad.
\label{disperion relation}
\end{equation}

If we instead vary the Lagrangian with respect to $\varphi$ we obtain an expression for $\mathcal{J}$:

\begin{equation}
\mathcal{J} [A] =  \frac{1}{\sqrt{2} \pi} \oint F^{\prime \prime} (q) [A - F(q)]^{1/2} d q \quad ,
\end{equation}
from this we obtain the equation for the conserved flux

\begin{equation}
\frac{a H^{\circ}_a}{\sqrt{2} \pi N} \oint F^{\prime \prime} (q) [A - F(q)]^{1/2} d q = \mathrm{const} \quad .
\label{flux conservation}
\end{equation}

An alternative form of the flux conservation equation can be obtained from the Lagrangian in $(\tilde{e},\tilde{e}_{\varphi})$ variables (Equation \ref{lagrangian in limit}). We can calculate $\mathcal{J}$ from this Lagrangian:

\begin{align}
\begin{split}
\mathcal{J} &=  \frac{\partial \langle \hat{\mathcal{L}} \rangle}{\partial k}\\
 &= \frac{\partial}{\partial k} \left( k^{-2} \frac{\tilde{e}^2}{2} - F(\tilde{e}_{\varphi}) \right) \\
&=  -k^{-3} \langle \tilde{e}^2 \rangle \quad ,
\end{split}
\end{align}
and obtain an equivalent conserved flux

\begin{equation}
\frac{a H_{a}^{\circ} N^2}{a \varphi_a} \langle e^2 \rangle = \mathrm{const} \quad .
\end{equation}

Combining this equation with the dispersion relation we obtain

\begin{equation}
  \langle e^2 \rangle \propto \frac{(a H_{a}^{\circ} N)^{-1}}{\oint F^{\prime \prime} (q) [A - F(q)]^{-1/2} d q } \quad .
\label{ecc env}
\end{equation}

\subsection{Expressions for eccentricity and nonlinearity}

The integral $\oint \cdot \, d q$ over the cycle of $q$ can be rewritten as $4 \int_{0}^{q_{+}} \cdot \, d q$. As such we can rewrite these integrals using Equation (\ref{complete untwisted I}) and (\ref{complete untwisted J}). The dispersion relation (Equation \ref{disperion relation}) becomes

\begin{equation}
a \varphi_a = \frac{\pi N}{\sqrt{2} I (q_{+})} \quad ,
\label{dispersion relation llscl}
\end{equation}
the conserved flux (Equation \ref{flux conservation}),

\begin{equation}
\frac{2 \sqrt{2} a H^{\circ}_a}{\pi N} J (q_{+}) = \mathrm{const} \quad ,
\label{conserved flux untwisted}
\end{equation}
and the equation for the eccentric envelope (Equation \ref{ecc env}) becomes

\begin{equation}
\langle e^2 \rangle \propto \frac{1}{a H_{a}^{\circ} N I (q_{+})} \quad .
\label{untwisted amplitude}
\end{equation}

In the linear limit (using $F = F^{\rm lin}$) we can evaluate $I$ and $J$ and obtain

\begin{equation}
I (q_{+})  = \frac{\pi}{2} \sqrt{\frac{2 \gamma - 1}{\gamma}} \quad ,
\end{equation}

\begin{equation}
J (q_{+}) = \frac{\pi}{16} \left(\frac{2 \gamma - 1}{\gamma} \right)^{3/2} q_{+}^{2} \quad .
\end{equation}

In the linear limit the solution to the short lengthscale oscillator has the form $e = e_0 \cos \varphi$, this gives a solution in the linear limit of

\begin{equation}
e \propto  (a H_{a}^{\circ} N)^{-1/2} \cos \left( \int \sqrt{\frac{2 \gamma}{2 \gamma - 1}}  \frac{N}{a} \, d a\right) \quad .
\label{lin sol}
\end{equation}

The equations for a 3D adiabatic eccentric disc in the linear limit, where the solution is assumed to vary on a length scale much shorter than $a$ is \citep{Teyssandier16}

\begin{equation}
 \Sigma^{\circ} a^3 n (\omega_f - \omega) e  =- \frac{\partial}{\partial a} \left( \frac{2 \gamma - 1}{2 \gamma} P^{\circ} a^3 e_a \right) \quad ,
\end{equation}
which can be rewritten in terms of $N$ and $a H^{\circ}_a$,

\begin{equation}
 H^{\circ}_{a} N^2 e  =- \frac{\partial}{\partial a}  \left( \frac{2 \gamma - 1}{2 \gamma} a  H^{\circ}_{a} a e_{a} \right) \quad .
\end{equation}
An approximate solution which can be obtained via WKB methods will yield equation (\ref{lin sol}), showing that the WKB limit of linear theory and the linear limit of the short wavelength theory for eccentric discs are the same.

\section{Conditions for Growth of Nonlinearity and Eccentricity} \label{untwisted cond}

We now turn to the question of under what circumstances do nonlinearity $q$ and eccentricity get amplified by the large scale disc properties. In effect, how does the disc behaviour control the long length scale envelope of the solution. Can a small disturbance in the outer region of the disc be focused and become highly nonlinear in the inner region? 
 
We show in Appendix \ref{integral proofs} that the integrals $I (q_{+})$, $J (q_{+}) $ are increasing functions of $q_{+}$. Using equation (\ref{conserved flux untwisted}) we can obtain a condition on the amplification of $q_{+}$ with decreasing $a$,

\begin{equation}
 \frac{\partial q_{+}}{\partial a} < 0 \quad \textrm{if and only if} \quad \frac{\partial}{\partial a} \left( \frac{a H^{\circ}_a}{N} \right) > 0
\end{equation}

The condition on $\langle e^2 \rangle$ to increase can be obtained from equation (\ref{untwisted amplitude}). From which we obtain the condition

\begin{equation}
\frac{\partial}{\partial a} \left( a H^{\circ}_a N \langle e^2 \rangle \right) < 0 \quad  \textrm{if and only if}  \quad \frac{\partial q_{+}}{\partial a} > 0 \quad .
\end{equation}
Similarly we have a condition on $a \varphi_a$

\begin{equation}
\frac{\partial}{\partial a}  \left( \frac{a \varphi_a}{N} \right) > 0 \quad \textrm{if and only if}  \quad \frac{\partial q_{+}}{\partial a} > 0 \quad .
\end{equation}

Using equations \ref{dispersion relation llscl} and \ref{untwisted amplitude} we can show that

\begin{equation}
\frac{a \varphi_a}{a \varphi_a |_{\rm lin}} = \frac{I(0)}{I(q)} \le 1 \quad ,
\end{equation}

\begin{equation}
\frac{\langle e^2 \rangle}{\langle e^2_{\rm lin} \rangle} = \frac{I(0)}{I(q)} \le 1 \quad ,
\end{equation}
where the subscript $_{\rm lin}$ denotes the prediction from linear theory. So in general we expect the nonlinear eccentric waves to behave like the linear eccentric waves, however amplification of the eccentricity and wavenumber of the eccentric wave due to wave focusing will be suppressed due to nonlinear effects. As such we can conclude that the wavenumber and (absolute) eccentricity of a nonlinear eccentric wave is bounded from above by that obtained from linear theory in the direction in which $q_{+}$ is increasing. 

Physically this can be understood as the nonlinear effect enhancing the pressure forces from the eccentric wave in the disc so that the eccentric wave does not need to attain as high an amplitude/wavenumber as in linear theory in order to balance the precessional effects captured by $\omega_{f} - \omega$.

If we now consider a disc with powerlaw profiles for $P^{\circ}$, $\Sigma^{\circ}$ and $\omega_{f} - \omega$; with $P^{\circ} \propto a^{r}$, $\Sigma^{\circ} \propto a^{\sigma}$ and $\omega_{f} - \omega \propto a^{s}$. The last of these corresponds to the non-Keplerian forces in the disc, in particular the lowest order contribution from GR is given by $s = -5/2$ and from a quadrupole term (such as that present in discs around binaries or rotating stars) $s = -7/2$. 

With these forms for the disc dependent quantities the linear theory predict that $q_{+}|_{\rm lin} \propto a^{\lambda}$ and $\langle e^2_{\rm lin} \rangle \propto a^{\mu}$ where the power law exponents are given by

\begin{equation}
\lambda = \frac{1}{4} \left( s + \sigma - 3 r - \frac{7}{2} \right) \quad ,
\end{equation}

\begin{equation}
\mu = -\frac{1}{2} \left(r + \sigma + s + \frac{9}{2} \right) \quad .
\end{equation}

For the nonlinear theory we obtain the following conditions for the behaviour of the eccentric mode,

\begin{equation}
 \frac{\partial q_{+}}{\partial a} < 0 \quad \textrm{if and only if} \quad \lambda < 0 \quad ,
\end{equation}
which is the same as that from linear WKB theory. The conditions on the behaviour of $\langle e^2 \rangle$ are

\begin{equation}
 \lambda > 0  \quad \mathrm{and} \quad \mu < 0 \implies \frac{\partial}{\partial a} \langle e^2 \rangle < 0 \quad ,
\label{plaw e2 env 1}
\end{equation}

\begin{equation}
 \lambda < 0  \quad \mathrm{and} \quad \mu > 0 \implies \frac{\partial}{\partial a} \langle e^2 \rangle > 0 \quad .
\label{plaw e2 env 2}
\end{equation}
These differ subtly from the condition derived from linear theory which is

\begin{equation}
\frac{\partial}{\partial a} \langle e^2 \rangle < 0 \quad \textrm{if and only if} \quad \mu < 0 \quad .
\end{equation}
This difference comes about because in nonlinear theory the sign of $\lambda$ controls whether nonlinear effect will compete with or enhance the forcing terms present in linear theory. In addition to conditions (\ref{plaw e2 env 1}) and (\ref{plaw e2 env 2}) there are two other possible combinations of signs for $\lambda$ and $\mu$.

\begin{enumerate}
\item $\lambda < 0  \quad \mathrm{and} \quad \mu < 0 \implies \frac{\partial}{\partial a} \langle e^2 \rangle$ bounded from below by linear theory.

\item $\lambda > 0  \quad \mathrm{and} \quad \mu > 0 \implies \frac{\partial}{\partial a} \langle e^2 \rangle$ bounded from above by linear theory.
\end{enumerate}

In both these cases we can make no definite conclusions about the sign of $\frac{\partial}{\partial a} \langle e^2 \rangle$ as the nonlinear effects compete with the linear terms. Instead $\frac{\partial}{\partial a} \langle e^2 \rangle$ has the same sign as linear theory for low $q_{+}$ and changes sign as $q_{+}$ increases.

In most regions of the disc (apart from the disc inner edge/gap where the disc is truncated) we expect $r \le 0$ and $\sigma \le 0$. In addition we expect the steepest precessional frequency that can be dominant in the disc to be that from a quadrupole term so $s \ge -7/2$.

\section{Modulation on the Disc Lengthscale in Power Law Discs} \label{results}

To illustrate our theory we calculate how the long length scale properties of an eccentric wave vary in a power law disc. 

\begin{figure}
\includegraphics[width=\linewidth]{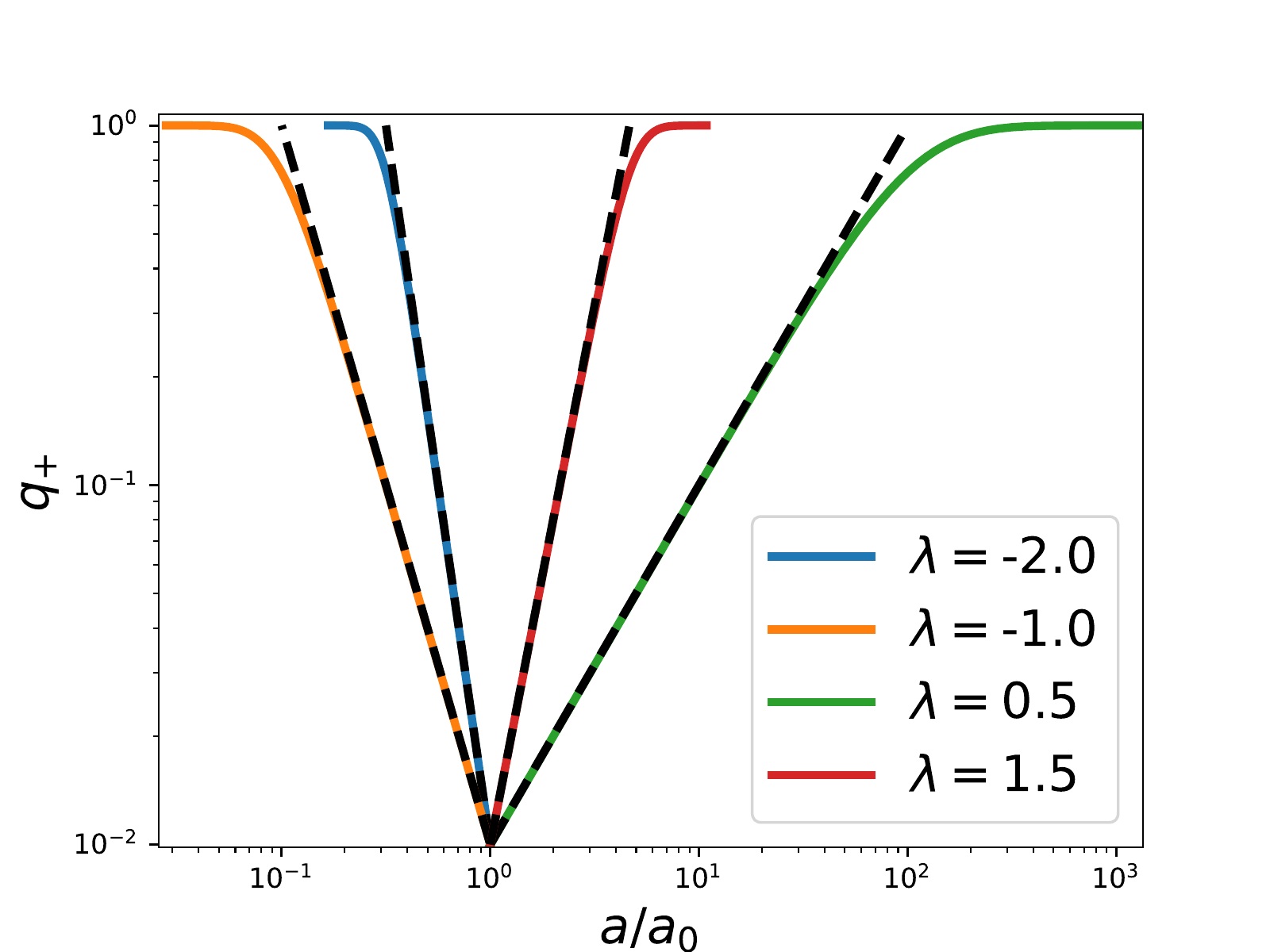}
\caption{Variation of the maximum nonlinearity attained during a wave cycle ($q_{+}$) on the disc lengthscale. Each solution is labelled by $\lambda$ which corresponds to the power law for $q_{+}$ in the linear limit. Dashed lines correspond to power law solutions from linear theory.}
\label{q profile}
\end{figure}

Figure \ref{q profile} shows how $q_{+}$ varies on the disc length scale dependent on $\lambda$. As expected the solution is a power law at low $q_{+}$ where linear theory is valid. As the nonlinearity increases the solution turns over due to the increasingly strong pressure forces which are able to more effectively balance the precessional forces in the nonlinear regime so that $q_{+}$ need not be as high as would be predicted in linear theory. The solution asymptotes to  $q_{+} \rightarrow 1$ as the precessional forces diverge which in a power law disc can only occur at $a \rightarrow 0$ or $a \rightarrow \infty$ dependent on the sign of $\lambda$.

\begin{figure}
\includegraphics[width=\linewidth]{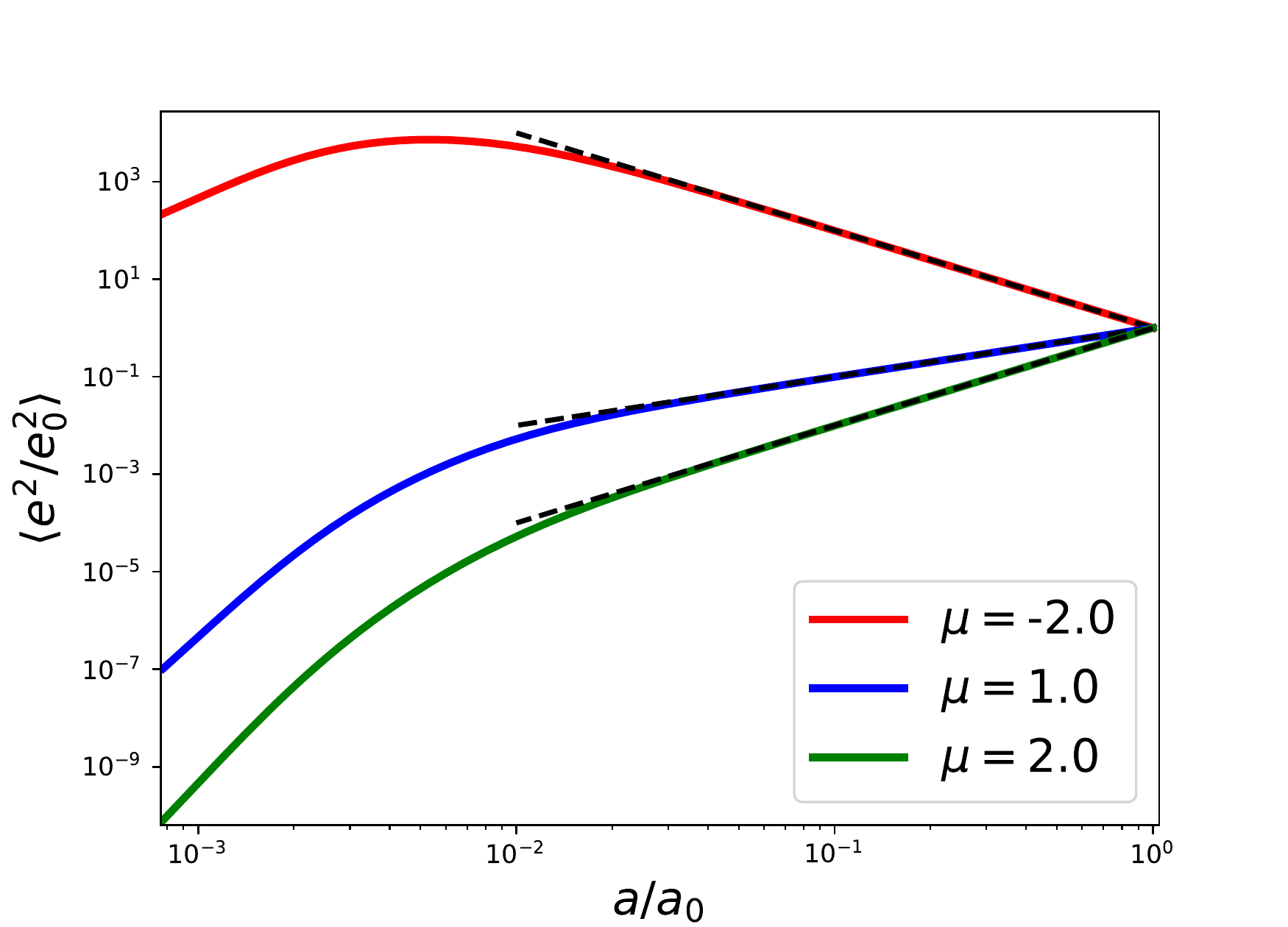}
\caption{Eccentric amplitude on the disc scale for $\lambda = -1$, as the theory is linear in eccentricity, the absolute value of this is unimportant. Dashed lines correspond to power law solutions from linear theory.}
\label{e profile neg}
\end{figure}

\begin{figure}
\includegraphics[width=\linewidth]{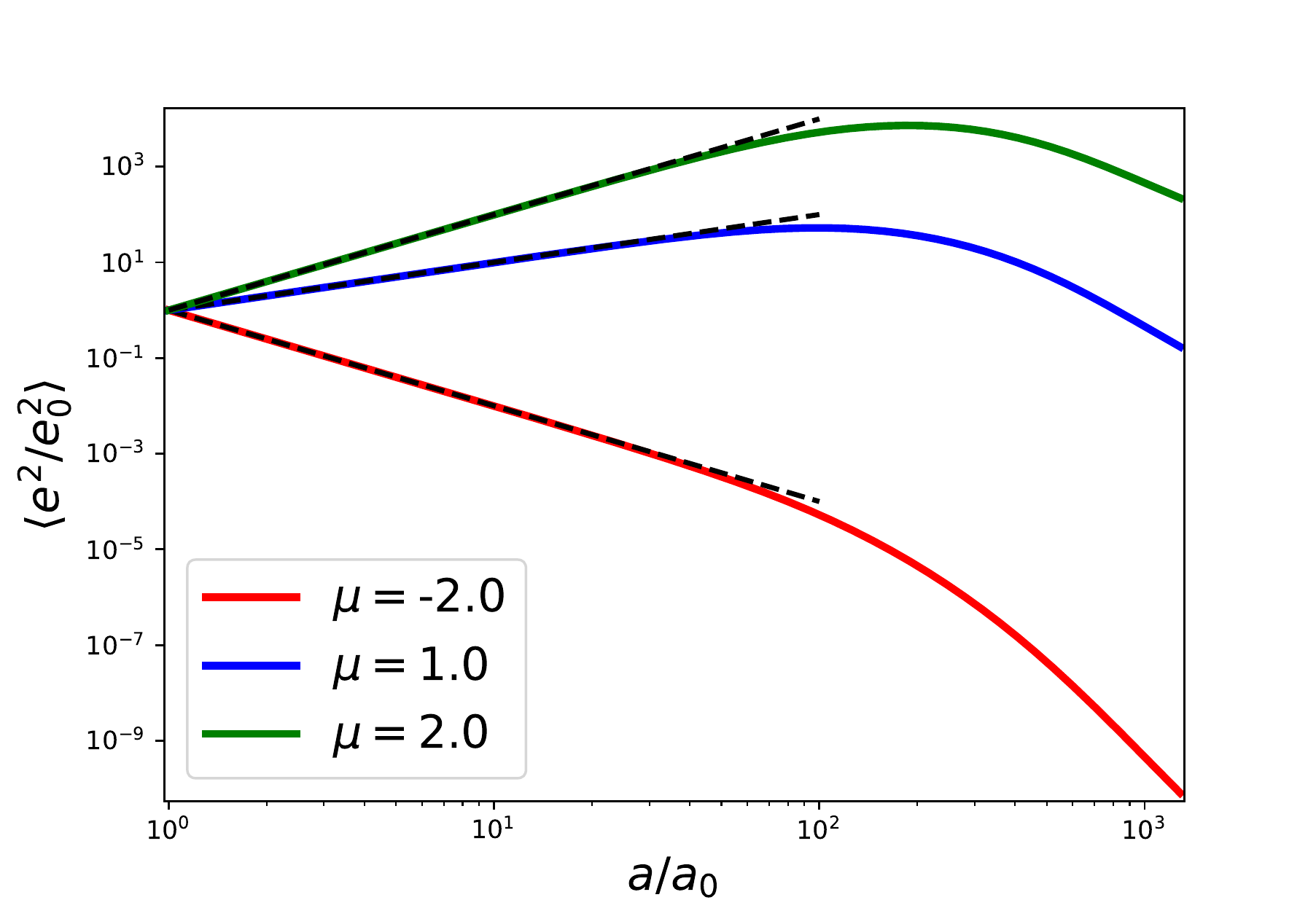}
\caption{Same as Figure \ref{q profile} but with $\lambda = 1$.}
\label{e profile pos}
\end{figure}

Figures \ref{e profile neg} and \ref{e profile pos} similarly show how $\langle e^2 \rangle$ varies on the disc length scale, with its behaviour dependent on both $\lambda$ and $\mu$. Unlike $q$ the equations are linear in $e$ and as such we are free to rescale our solutions by a constant value. As with $q_{+}$ the solution is a power law in the linear regime and turns over at increasing nonlinearities so that it always lies below the power law slope of linear theory.

\section{Discussion} \label{astro consequences}

\subsection{Exploring the parameter space relevant to astrophysical discs.}

Now we consider some astrophysical consequences of the conditions we have derived for eccentric wave steepening. To do so we will concentrate on the conditions derived for power law discs as these are sufficient to draw conclusions relevant to an astrophysical context. We will assume that the power law profile for the surface density is bounded such that $-2 < \sigma < 0$, which simply states that the surface density must decrease outwards but most of the mass of the disc is at large radii.

The local precessional terms $\omega_{f} - \omega$ are assumed to be dominated by one precessional effect, and can thus be described by a power law. As above we use $\omega_{f} - \omega \propto a^s$. Different regions of the disc are likely to be dominated by different precessional terms and the eccentric mode will switch between them as it moves to smaller semimajor axis. In general $s$ should decrease moving from the outer to the inner disc as more centrally concentrated forces become important. Starting from precessional effects that dominate the outer disc and moving inwards, some important precessional effects and their associated values for $s$ are

\begin{itemize}
\item Outer quadrupole from an external perturber: $s = 3/2$.
\item Global (retrograde) disc precession: $s = 0$.
\item Breathing mode: The effect of the breathing mode can be approximated with $s = r - \sigma - 1/2$. This is a special case of the precessional effects as it depends on the pressure profile. More generally the effect of the breathing mode should be included in the  functional form of $F$.
\item General relativity: The lowest order GR correction has $s = -5/2$.
\item Inner quadrupole term: $s = -7/2$.
\end{itemize} 

\begin{figure}
\includegraphics[width=\linewidth]{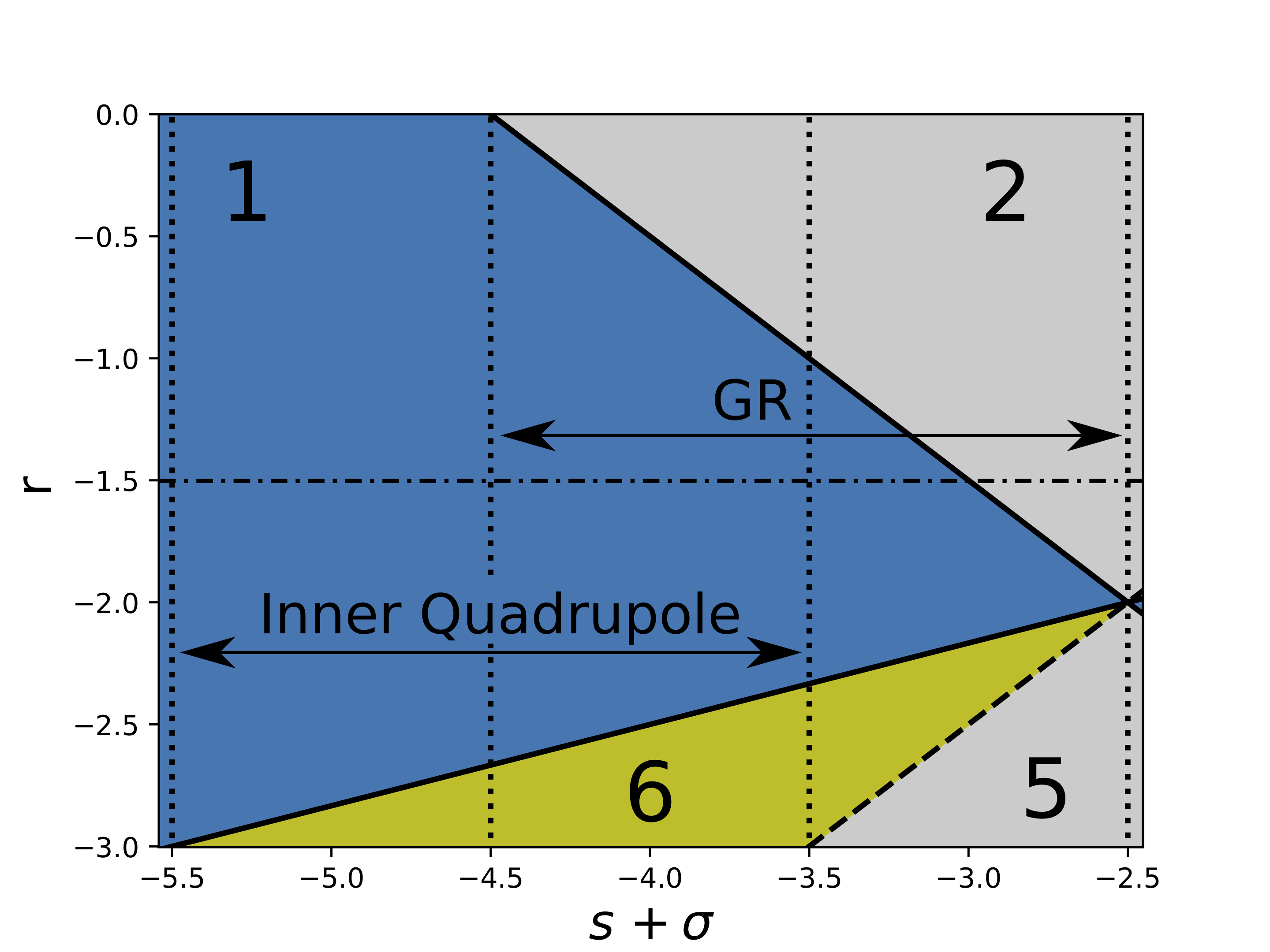}
\caption{Different regimes (see text) for eccentric modes for the inner region of discs with power law profiles. The solid lines separate regions which dictate the behaviour of the eccentric mode. The dashed line corresponds to the precession frequency associated with pressure forces (such as the breathing mode), thus here $\partial_a N = 0$ as the precessional forces are proportional to the pressure forces here. The dotted lines show the region of the parameter space occupied by discs with a given dominant precessional effect assuming the power law for the surface density profile is in the range $-2 < \sigma < 0$. In general the dominant precessional effect changes in a disc and moves towards the left of the diagram at smaller radii, with the dash-dotted line corresponding to the pressure profile of a steady $\alpha$-disc.}
\label{eccentric param space inner}
\end{figure}

\begin{figure}
\includegraphics[width=\linewidth]{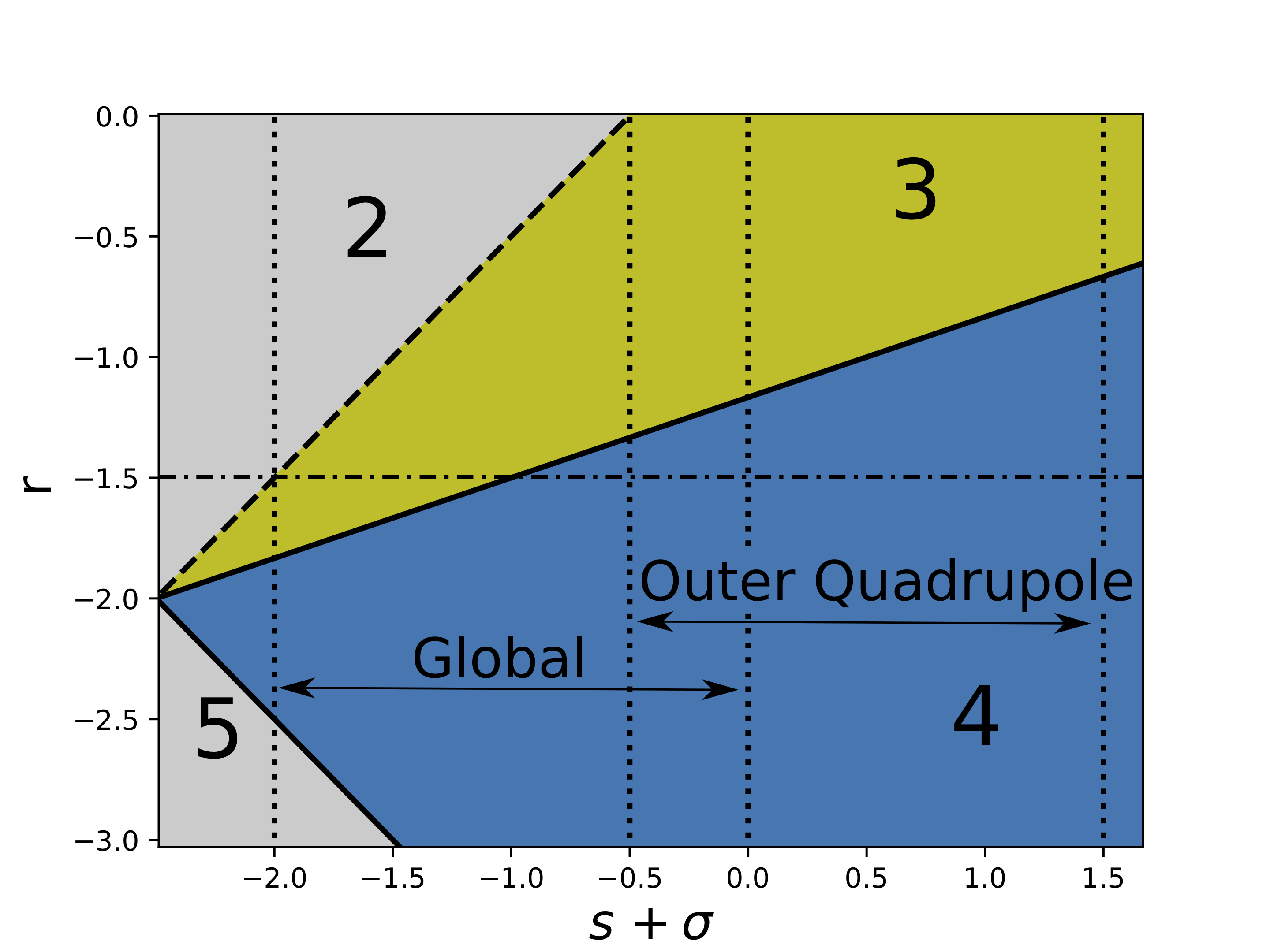}
\caption{Same as Figure \ref{eccentric param space inner} but for the outer part of the disc. The precessional forces that are important here are those which increase outwards (for instance those resulting from an external perturber), as only these will compete with the global disc precession at large radii.}
\label{eccentric param space outer}
\end{figure}

Figures \ref{eccentric param space inner} and \ref{eccentric param space outer} shows how these different precessional regimes and surface density/pressure profiles affect the behaviour of the eccentric mode. It should be noted that most discs will not occupy a single point on these diagrams but will slowly move left to smaller $s+\sigma$ at smaller radii as short range forces become important. We consider vertically integrated pressure profiles $P^{\circ} \propto a^{r}$ with $-3 < r < 0$. A steady $\alpha$-disc should have $r=-1.5$.

The different regions of eccentric mode behaviour are

\begin{enumerate}
\item $\frac{\partial q_{+}}{\partial a} < 0$, $\frac{\partial}{\partial a} N^2 < 0$, $\frac{\partial}{\partial a} \langle e^2 \rangle > 0$.
\item $\frac{\partial q_{+}}{\partial a} < 0$,  $\frac{\partial}{\partial a} N^2 < 0$, $\frac{\partial}{\partial a} \langle e^2 \rangle$ < 0 for small $q_{+}$, $\frac{\partial}{\partial a} \langle e^2 \rangle$ > 0 for large $q_{+}$.
\item $\frac{\partial q_{+}}{\partial a} < 0$, $\frac{\partial}{\partial a} N^2 > 0$,  $\frac{\partial}{\partial a} \langle e^2 \rangle$ < 0 for small $q_{+}$, $\frac{\partial}{\partial a} \langle e^2 \rangle$ > 0 for large $q_{+}$.
\item $\frac{\partial q_{+}}{\partial a} > 0$, $\frac{\partial}{\partial a} N^2 > 0$, $\frac{\partial}{\partial a} \langle e^2 \rangle < 0$.
\item $\frac{\partial q_{+}}{\partial a} > 0$, $\frac{\partial}{\partial a} N^2 > 0$, $\frac{\partial}{\partial a} \langle e^2 \rangle$ > 0 for small $q_{+}$, $\frac{\partial}{\partial a} \langle e^2 \rangle$ < 0 for large $q_{+}$. This region rarely occurs in real discs as this requires a particularly steep pressure gradient.
\item $\frac{\partial q_{+}}{\partial a} > 0$, $\frac{\partial}{\partial a} N^2 < 0$, $\frac{\partial}{\partial a} \langle e^2 \rangle$ > 0 for small $q_{+}$, $\frac{\partial}{\partial a} \langle e^2 \rangle$ < 0 for large $q_{+}$.  As with region 5, this region rarely occurs in real discs.
\end{enumerate}

Regions $1-4$ all feature focusing of the eccentric wave (i.e. an increase in either $q_{+}$ or $\langle e^2 \rangle$ inwards) by the background disc. Whereas regions $5$ and $6$ (where the pressure gradients are very steep) are defocusing. In regions $2$ and $3$ the essentially linear phenomena of wave focusing competes against the nonlinear enhancement of pressure forces in the wave, resulting in a change in behaviour when the wave is sufficiently nonlinear.

The inner part of the disc is covered by Figure \ref{eccentric param space inner} and includes regions $1,2,5,6$. As $N^2$ needs to be large in order that the separation of scales in the short wavelength limit remains valid, the dashed line corresponding to the breathing mode (and $\partial_a N^2 = 0$) is important in determining the validity of the theory. Considering an eccentric wave that starts in the short wavelength limit in the outer regions, then we require $\partial_a N^2 < 0$ for the theory to be applicable in the inner disc.  Only region 5 violates this condition, however this region is unlikely to be important for real discs due to the steep pressure gradients required.

The outer part is covered by Figure \ref{eccentric param space outer} and includes regions $2-5$. Starting with an eccentric wave in the inner region in the short wavelength limit we require  $\partial_a N^2 > 0$ for the theory to be applicable to the outer disc. Region 2 violates this; unlike region 5 in the inner disc, region 2 can be important to the outer part of discs with retrograde precession and shallow pressure gradients.

The reader might be concerned with the fact that a disc with inner quadrupole precession sits entirely in region 1, which predicts that $\frac{\partial}{\partial a} \langle e^2 \rangle > 0$. This contradicts the results of many hydrodynamic simulations of circumbinary discs (e.g. \citet{MacFadyen08}, \citet{Miranda17}, \citet{Thun17}, \citet{Thun18}, \citet{Miranda18}). It should be noted however that this result is only true for a wavelike eccentric mode. The modes set up in simulations are better thought of as confined modes trapped by the quadrupole potential and are non-oscillatory so the theory developed here is not applicable. It is possible that in the outer disc far from the binary where the evanescent wave has declined to zero that the wavelike mode should dominate. However this will require a mechanism for a binary to excite such a wave and maintain it against dissipation. As the confined and wavelike modes are nonlinear in different parts of the disc we expect that these two modes can be superposed with each dominating the flow geometry in a particular part of the disc.
 
\subsection{Eccentric  waves in an illustrative astrophysical system}

An astrophysical system which highlights the role of the different precessional effects on the behaviour of the eccentric discs are X-ray binaries, which was studied by \citet{Ferreira09} using linear theory. While it should be cautioned that the theory we have developed is not directly applicable to XRBs due to dissipative effects from MRI turbulence which is neglected, studying the ideal theory in such a system provides insight into the nonlinear behaviour of eccentric waves in real discs.

 If we assume that a linear eccentric wave is generated in the outer region of the disc, most likely due to excitation at the 3:1 Lindblad resonance \citep{Lubow91,Lubow91b}, we can follow this eccentric wave through the different regions of the disc where different forces  dominate the precessional effects to see how it develops in the inner disc.

In the outermost part of the disc the dominant precessional force is from the external companion. Here the the disc will be in either region 3 or 4, for $r=-1.5$ (typical pressure exponent for an $\alpha$-disc) the disc will be in region 4 and we expect the nonlinearity of the wave to decrease as it propagates inwards while its eccentricity increases.

If this precessional term dominates over a large radial extent in the disc then the eccentric wave will exit the short-wavelength regime and the theory developed is no longer valid. In this case we would expect that the eccentric wave will be global in character and vary on the disc lengthscale. However as the eccentric wave propagates away from the disc edge other precessional terms become important. The next precessional effect to occur is a global (retrograde) precession which results in the disc sitting in region 3 or 4. Here we expect the nonlinearity to start increasing again while the increase in the eccentricity tails off. The breathing mode can also be important here with a similar effect to a global eccentric mode; however its effects can extend into the inner disc extending region 2/3 further in than might be expected from 2D theory.

In the inner disc general relativistic precession will start to compete with the breathing mode, resulting in the disc transitioning into regions 1 or 2. This will cause the nonlinearity to increase while the amplification of the eccentricity will stall and eventually decline. In general regions 1,2 are conducive to the formation of a nonlinear short-wavelength mode. As such, if the eccentricity of the eccentric wave generated by the external companion has not increased to too high a value in the outer disc, we expect a short-wavelength mode as studied here to form in the inner disc (this can be regarded as a nonlinear extension of the results of \citet{Ferreira09}). The last region that we have left off is the internal quadrupole; while present due to higher order GR correction, this is less important to the eccentric wave as it never dominates over the lower order GR precession considered until the marginally stable orbit is reached. 

Near the marginally stable orbit the pressure and surface density profiles turn over so that they decline inward. As GR precession will still be dominant in this region this will force the solution into region 2 resulting in the nonlinearity increasing inwards and possible increase of the eccentricity. In the linear theory \citep{Ferreira09} the eccentricity and nonlinearity diverge at the disc edge. Here we predict that the eccentricity can grow, but must do so more slowly than the linear predictions of \citet{Ferreira09}, however this may still be enough for the short-wavelength theory to predict its own breakdown in this region. Unlike \citet{Ferreira09} the absolute value of the nonlinearity has physical significance and cannot exceed $q=1$. 

\subsection{Towards an extension of the results of \citet{Ferreira09}}

Following \citet{Ferreira09} we consider the behaviour of an eccentric disc in a steady alpha model \citep{Shakura73}; however we neglect the effects of dissipation on the eccentric mode. We make use of the relativistic expressions for the characteristic frequencies in a Kerr metric \citep{Kato90}

\begin{equation}
\Omega = (r^{3/2} + s)^{-1} \quad ,
\end{equation}

\begin{equation}
\kappa = \Omega \sqrt{1 - \frac{6}{r} + \frac{8 s}{r^{3/2}} - \frac{3 s^2}{r^2}} \quad ,
\end{equation}
where in this section $r$ is the cylindrical radius in units of $GM/c^2$, $s$ is the dimensionless spin parameter of the black hole $(-1<s<1)$ and the frequencies are in units of $c^3/GM$. In the Shakura \& Sunyaev model the disc is split into an inner radiation pressure dominated region (region a) and an outer gas pressure dominated region (region b). In region a)  \citep{Ferreira09},

\begin{equation}
\Sigma^{\circ} \propto r^{3/2} f^{-1} \quad ,
\end{equation}

\begin{equation}
H \propto f \quad ,
\end{equation}
while in region b)

\begin{equation}
\Sigma^{\circ} \propto r^{-3/5} f^{3/5} \quad ,
\end{equation}

\begin{equation}
H \propto r^{21/20} f^{1/5} \quad ,
\end{equation}
where we have introduced $f = 1 - \sqrt{r_{in}/r}$ with $r_{in}$ the dimensionless, spin dependent, radius of the marginally stable orbit. 

This allows us to obtain the background profiles $P^{\circ}$, $\Sigma^{\circ}$, $\omega_f$ in the unperturbed circular disc. As noted in \citet{Ferreira09} this background disc model is strictly Newtonian, apart from the truncation of the inner edge at the marginally stable orbit. The most important relativistic effect on the eccentric mode is the contribution to apsidal precession.

Making the approximation $r \approx a$ which is correct in the limit $N^2 \gg 1$ we obtain expressions for $a H^{\circ}_a$ and $N$ which can be used to determine the eccentric disc dynamics:

\begin{equation}
a H^{\circ}_a \propto a^2 P^{\circ} \propto a^{1/2} f  \quad ,
\end{equation}

\begin{equation}
N^2 \propto \frac{\Sigma^{\circ}}{a H_a^{\circ}} a^{-1}( 6 a - 8 s a^{1/2} + 3 s^2 )  \quad .
\end{equation}

The behaviour of $N$ depends on whether the disc is gas or radiation pressure dominated. 

a) Radiation pressure dominated

\begin{equation}
N^2 \propto f^{-2} ( 6 a - 8 s a^{1/2} + 3 s^2 ) \quad ,
\end{equation}

b) Gas pressure dominated

\begin{equation}
N^2 \propto f^{-2/5} a^{-21/10} ( 6 a - 8 s a^{1/2} + 3 s^2 ) \quad .
\end{equation}

In the outer part of the disc (region b of the Shakura \& Sunyaev model) our theory predicts that the eccentric wave should be in region 1 where eccentricity decreases inwards but nonlinearity and $N^2$ increase. This is the same behaviour as found in \citet{Ferreira09} as the nonlinearity doesn't modify the qualitative behaviour in this region. The main effect of nonlinearity will be to make the inward decrease of the eccentricity stronger so that for a given radius the eccentricity in the nonlinear theory will be lower than that predicted in \citet{Ferreira09}.

Consider the region close to the marginally stable orbit and assume that this corresponds to region a) of the Shakura \& Sunyaev model. In the linear regime the envelopes for $q$ and $\langle e^2 \rangle$ are given by

\begin{equation}
q \propto f^{-1} a^{-1/4} (6 a - 8 s a^{1/2} +3 s^2)^{1/4} \quad ,
\end{equation}

\begin{equation}
\langle e^2 \rangle \propto a^{-1/2} (6 a - 8 s a^{1/2} +3 s^2)^{-1/2}  \quad ,
\end{equation}
which shows a divergence of $q$ as the wave approaches the marginally stable orbit, while the eccentricity should reach a finite value. In nonlinear theory in this part of the disc we obtain the following for $J(q_{+})$:

\begin{equation}
J(q_{+}) \propto f^{-2} a^{-1/2} (6 a - 8 s a^{1/2} +3 s^2)^{1/2}
\end{equation}

Near the marginally stable orbit $f \rightarrow 1$ so we require that $J(q_{+})$ diverges. In linear theory as $J(q_{+}) \propto q_{+}^2$ this would cause $q_{+}$ to diverge. However in nonlinear theory $J(q_{+})$ diverges when $q_{+} \rightarrow 1$ so we can conclude that $q_{+} \rightarrow 1$ as $f \rightarrow 0$. For the eccentricity we can write

\begin{equation}
\langle e^2 \rangle = \langle e^{2}_{\rm lin} \rangle \frac{I(0)}{I(q_{+})}
\end{equation}

As $\langle e^{2}_{\rm lin} \rangle$ tends to a constant as $f \rightarrow 0$ and $I(q_{+})$ diverges as $q_{+} \rightarrow 1$ we find that $\langle e^2 \rangle \rightarrow 0$ as $f \rightarrow 0$

Given these results, the expected behaviour for an eccentric wave approaching the marginally stable orbit is for the wave to circularise while approaching an orbital intersection. This may have important consequences for the boundary conditions applicable at this location, as the near orbital intersection at the marginally stable orbit suggests that the wave should be highly dissipated here.

Eccentric waves in the inner region of XRBs are of interest as a driving mechanism for high frequency quasi-periodic oscillations (QPOs), a type of variability observed in the X-ray spectra of some black hole systems, via the excitation of trapped inertial waves \citep{Kato08,Ferreira08,Ferreira09,Dewberry18,Dewberry19}. It appears from our work that the short-wavelength limit of eccentric waves is important for the inner regions of XRBs, which is a nonlinear extension of those considered by \citet{Ferreira08}, \citet{Ferreira09} in this context. \citet{Ferreira08} showed that the growth rate of trapped inertial waves from interaction with the eccentric background is proportional to $|r \frac{d E}{d r}|^2 \approx q^2$, as such nonlinear eccentric waves should couple more strongly to inertial waves and thus be more effective at generating QPOs . If this short-wavelength theory is to be useful for describing the QPO generation mechanism, then it must be extended to include the effect of dissipation due to the strong  MRI turbulence expected in XRBs - along with the potentially significant damping of the mode when highly nonlinear. We leaves this extension to future work.

\section{Conclusion} \label{conc}

We have formulated a nonlinear theory for short wavelength eccentric waves in astrophysical discs based on the averaged Lagrangian method of \citet{Whitham65}. We have found that the linear theory can be used to provide bounds on the nonlinear behaviour and can thus be useful in determining the large lengthscale behaviour. We have also derived conditions under which nonlinearity and/or eccentricity can grow through wave steepening in a disc and discussed what effect this has on eccentric waves in various astrophysical settings.

In general we have found that the inclusion of the nonlinearity has the following effects.

\begin{itemize}
\item There are conditions under which eccentric waves remain in the short-wavelength limit but become increasingly nonlinear as they propagate inwards. These depend on the precessional effects in the disc and the surface density/pressure profiles. This is a form of wave-focusing of the eccentric waves which can occur in astrophysical discs. While predicted in linear theory, we have shown that this wave focusing occurs even in the presence of strong nonlinearity. 
\item  There exists an highly nonlinear limit where the eccentric wave becomes a triangle wave and the maximum nonlinearity in a wave cycle $q_{+}$ approaches $1$. In this limit the eccentric wave oscillates between nearly intersecting orbits, with alternating argument of pericentre, on a short radial lengthscale comparable to (but slightly longer than) the scale height.
\item Nonlinearity prevents the generation of an orbital intersection by strengthening the pressure forces so that it is necessary for the precessional forces to diverge in order that an orbital intersection is generated. 
\end{itemize}

The interior region of an X-ray binary should naturally support such short-wavelength eccentric waves, while the orbiting companion provides a mechanism by which they can be generated. These eccentric waves are of interest as a driving mechanism for QPOs. Future work should determine the effect of dissipation on these waves, which has been shown to be important in linear theory.

\section*{Acknowledgements}

E. Lynch would like to thank the Science and Technologies Facilities Council (STFC) for funding this work through a STFC studentship. This research was supported by STFC through the grant ST/P000673/1.




\bibliographystyle{mnras}
\bibliography{FocusingofNonlinearEccentricWaves.bib} 



\appendix

\onecolumn

\section{Hamiltonian in the Short Wavelength Limit} \label{geo F deriv}

As shown by \citet{Ogilvie19} the Hamiltonian for a disc of ideal fluid with only Keplerian and pressure forces can be written as

\begin{equation}
H = \int H^{\circ}_a F d a \quad,
\end{equation}
 where $H^{\circ}_a$ is the Hamiltonian density of the equivalent circular disc, and $F$ is a geometric factor that depends on the geometry of the orbits and the ratio of specific heats $\gamma$. The general form of $F$ for different disc models is

\begin{equation}
F = \langle F_E (j,h) \rangle \quad ,
\end{equation}
where $j = j(e,q,\cos(E - \alpha))$ is a dimensionless Jacobian with $j= 1$ for a circular disc; $h$ is a dimensionless scale-height (again with $h=1$ for a circular disc); and the angle brackets denote the orbit average

\begin{equation}
\langle \cdot \rangle := \frac{1}{2 \pi} \int_0^{2 \pi} \cdot \, (1 - e \cos E) \, d E \approx \frac{1}{2 \pi} \int_0^{2 \pi} \cdot \, d E + O(e) \quad .
\end{equation}

For the isothermal case it can be shown that $\langle \ln h(e,q,E - \alpha) \rangle = O(e^2)$ \citep{Ogilvie19} and as such this term can be neglected. For the adiabatic disc the equation governing the evolution of the non-dimensional scaleheight $h$ is

\begin{equation}
(1 - e \cos E) \frac{d^2 h}{d E^2} - e \sin E \frac{d h}{d E} + h = \frac{(1 - e \cos E)^3}{j^{\gamma - 1} h^{\gamma}} \quad .
\end{equation}
where $E$ is the eccentric anomaly. At lowest order in $e$ this becomes

\begin{equation}
 \frac{d^2 h}{d E^2} + h = \frac{1}{j^{\gamma - 1} h^{\gamma}} = \frac{1}{j(0,q, E - \alpha)^{\gamma - 1} h^{\gamma}} \quad ,
\end{equation}
so we conclude that the nondimensional scaleheight has a solution of the form $h(e,q,E - \alpha)  = h_0(q,E - \alpha)  + O(e)$. Using this in the functional form of $F$ given above, the lowest order terms in $e$ are

\begin{equation}
F = \langle F_E (j(e,q,E - \alpha),h(e,q,E - \alpha)) \rangle \approx \langle F_E (j(0,q,E),h(0,q,E)) \rangle + O(e)
\end{equation}
where we can remove the alpha dependence by a phase shift of the integration variables. Hence we conclude that these forms of the geometric part of the Hamiltonian can be written in the form $F = F(q)$.

In general if we have any F of the form

\begin{equation}
F = \langle F_{E} (e,q,E - \alpha) \rangle \quad ,
\end{equation}
then if $F_{E} (0,q,E - \alpha)$ is well defined (which must be the case in order that the disc has a circular limit) 

\begin{equation}
F (e,q,\alpha) = \langle F_{E} (0,q,E - \alpha) \rangle + O(e) =  \langle F_{E} (0,q,E) \rangle + O(e) \approx F(q) \quad .
\end{equation}

In other words the low amplitude limit of $F$, $\lim_{e \rightarrow 0} F (e,q,\alpha) \approx F(q)$ is valid for any disc with a circular limit and global rotation symmetry. 

\section{Behaviour of the Wave Action Integrals} \label{integral proofs}

In this Appendix we introduce various integrals of $F$ which are important in describing the nonlinear behaviour of the eccentric waves. We shall also prove a few results which are important for deriving conditions on the behaviour of the untwisted eccentric modes, given reasonable assumptions about the properties of $F$. Throughout this section we shall assume that $F$

\begin{enumerate}
\item Is a monotonically increasing function of $|q|$;
\item Has a monotonically increasing derivative (and thus $F^{\prime \prime} > 0$);
\item Diverges (with diverging derivative) as $|q| \rightarrow 1$.
\end{enumerate}

These are exactly the same assumptions on $F$ that are used to show that $q$ oscillates between two values $q_{+}$ and $-q_{+}$. Note $F^{\prime \prime} = 0$ is a singular point of the equation describing the short lengthscale behaviour (equation \ref{e eom}); this singular point separates oscillatory and evanescent behaviour of the solution on the short lengthscale. 

All given forms of $F$ considered obey these properties, and any physical forms of $F$ should diverge as $|q| \rightarrow 1$ due to the extreme pressure gradients associated with orbital intersection. If $F$ does not satisfy these properties (but does diverge at $|q| \rightarrow 1$) then it is possible that wave cannot exceed some critical $q$.

We make two additional assumptions about the properties of $F$ which are needed for some of the results in this section. These are that

\begin{enumerate}
\setcounter{enumi}{3}
\item $F^{\prime \prime}$ is a nondecreasing function of $|q|$ (or $q F^{\prime \prime \prime} (q) \ge 0$);
\item $F^{\prime \prime \prime \prime} (q) \ge 0$.
\end{enumerate}

These are additional conditions strengthening the convexity of $F$. Although reasonable, these are harder to justify than conditions $1-3$ which ought to be satisfied by any physically relevant forms of $F$ which produce oscillatory eccentric waves. Both $F^{(2 \rm D)}$ and $F^{(\rm iso)}$ satisfy conditions $5$ and $6$, and by a continuity argument 3D discs with $\gamma$ sufficiently close to 1 must remain close to the 2D (or isothermal) cases so that the conditions remain satisfied. It is possible that a sufficiently incompressible disc (or some other disc model not discussed here) might break one of these conditions, in which case the conclusions on the behaviour of eccentric waves in these discs as derived in this paper would have to be revisited for these disc models. The conditions we derive here are fairly conservative sufficiency conditions so even if a disc model breaks condition $5$ or $6$ the results we derive here may still hold.

We make use of various integrals of $F$ in order to describe the nonlinear behaviour of the eccentric wave. To describe the short lengthscale nonlinear oscillator we make use of

\begin{equation}
I (q,q_{+}) :=  \int_{0}^{q} F^{\prime \prime} (x) [F(q_{+}) - F(x)]^{-1/2} d x \quad ,
\label{incomplete untwisted I}
\end{equation}

\begin{equation}
J (q,q_{+}) :=  \int_{0}^{q} F^{\prime \prime} (x) [F(q_{+}) - F(x)]^{1/2} d x  \quad .
\end{equation}
and for the long lengthscale behaviour we define complete forms of these integrals

\begin{equation}
I (q_{+}) := I (q_{+},q_{+}) =  \int_{0}^{q_{+}} F^{\prime \prime} (q) [F(q_{+}) - F(q)]^{-1/2} d q \quad ,
\end{equation}

\begin{equation}
J (q_{+}) := J (q_{+},q_{+}) = \int_{0}^{q_{+}} F^{\prime \prime} (q) [F(q_{+}) - F(q)]^{1/2} d q \quad ,
\end{equation}

Given the assumptions about the properties of $F$, it can be seen that these integrals are strictly positive. Whether $I(q_{+})$ and $J(q_{+})$ are increasing or decreasing functions of $q_{+}$ is important in determining the behaviour of the untwisted eccentric modes on the disc lengthscale. 

We can show that, for $F$ satisfying the assumptions above, $\partial_{q_{+}} J (q_{+})  > 0$:

\begin{align}
\begin{split}
\partial_{q_{+}} J (q_{+}) &=  \int_{0}^{q_{+}} \frac{\partial}{\partial q_{+}} \left( F^{\prime \prime} (q) [F(q_{+}) - F(q)]^{1/2} \right) \, d q \\
&= \frac{1}{2} \int_{0}^{q_{+}} F^{\prime} (q_{+})  F^{\prime \prime} (q) [F(q_{+}) - F(q)]^{-1/2} \, d q + \frac{1}{2} F^{\prime} (q_{+})  F^{\prime \prime} (q) [F(q_{+}) - F(q)]^{-1/2} \Biggl|_{q = q_{+}} > 0 \quad ,
\end{split}
\end{align}
as the integrand is positive and the boundary term is zero. Showing a similar result for $I(q_{+})$ is less straightforward as differentiating $I(q_{+})$ using in the normal manner using the Leibniz rule results in a divergent integral along with a divergent contribution from the boundary. The way in which these cancel is not clear. As such we need to proceed more carefully. Consider an expansion of $I(q_{+} + x)$ with $x \ll 1$ and introducing a new renormalised variable $\tilde{q} = \frac{q_{+}}{q_{+} + x} q$ then we obtain the expansion

\begin{align}
\begin{split}
I(q_{+} + x) &=  \int_{0}^{q_{+} + x} F^{\prime \prime} (q) [F(q_{+} + x) - F(q)]^{-1/2} \, d q \\
&= \int_{0}^{q_{+}} F^{\prime \prime} (\frac{q_{+} + x}{q_{+}} \tilde{q}) [F(q_{+} + x) - F(\frac{q_{+} + x}{q_{+}} \tilde{q})]^{-1/2} \left( \frac{q_{+} + x}{q_{+}} \right) \, d \tilde{q} \\
&\approx I(q_{+}) + x \int^{q_{+}}_{0} \frac{2 [F(q_{+}) - F(\tilde{q})] [F^{\prime \prime} (\tilde{q}) + \tilde{q} F^{\prime \prime \prime} (\tilde{q})] -[q_{+} F^{\prime} (q_{+}) - \tilde{q}F^{\prime} (\tilde{q})] F^{\prime \prime}(\tilde{q}) }{2 q_{+} [F(q_{+}) - F(\tilde{q})]^{3/2}} d \tilde{q} \\
&+  x^2 \int^{q_{+}}_{0} \frac{1}{8 q^2_{+}  [F(q_{+}) - F(\tilde{q})]^{5/2} }  \Biggl( 3 (\tilde{q}  F^{\prime} (\tilde{q}))^2 F^{\prime \prime} (\tilde{q}) + 2 [F(q_{+}) - F(\tilde{q})] [\tilde{q}^2 F^{\prime \prime} (\tilde{q})  + 2 \tilde{q} F^{\prime} (\tilde{q}) ( F^{\prime \prime} (\tilde{q}) + \tilde{q} F^{\prime \prime \prime} (\tilde{q}))] \\
&+ 4 [F(q_{+}) - F(\tilde{q}) ]^2 [2 \tilde{q} F^{\prime \prime \prime} (\tilde{q}) + \tilde{q}^2 F^{\prime \prime \prime \prime} (\tilde{q})] \Biggr) d \tilde{q} + O(x^3)\quad ,
\end{split}
\end{align}
from which we obtain the derivative by $\partial_{q_{+}} I(q_{+}) = \partial_{x} I(q_{+} + x) |_{x=0}$

\begin{equation}
\partial_{q_{+}} I(q_{+}) = \int^{q_{+}}_{0} \frac{2 [F(q_{+}) - F(\tilde{q})] [F^{\prime \prime} (\tilde{q}) + \tilde{q} F^{\prime \prime \prime} (\tilde{q})] -[q_{+} F^{\prime} (q_{+}) - \tilde{q}F^{\prime} (\tilde{q})] F^{\prime \prime}(\tilde{q}) }{2 q_{+} [F(q_{+}) - F(\tilde{q})]^{3/2}} d \tilde{q} \quad ,
\end{equation}
and second derivative $\partial^2_{q_{+}} I(q_{+}) = \partial^2_{x} I(q_{+} + x) |_{x=0}$

\begin{align}
\begin{split}
\partial^2_{q_{+}} I(q_{+}) &=  \int^{q_{+}}_{0} \frac{1}{4 q^2_{+}  [F(q_{+}) - F(\tilde{q})]^{5/2} }  \Biggl( 3 (\tilde{q}  F^{\prime} (\tilde{q}))^2 F^{\prime \prime} (\tilde{q}) + 2 [F(q_{+}) - F(\tilde{q})] [\tilde{q}^2 F^{\prime \prime} (\tilde{q})  + 2 \tilde{q} F^{\prime} (\tilde{q}) ( F^{\prime \prime} (\tilde{q}) + \tilde{q} F^{\prime \prime \prime} (\tilde{q}))] \\
&+ 4 [F(q_{+}) - F(\tilde{q}) ]^2 [2 \tilde{q} F^{\prime \prime \prime} (\tilde{q}) + \tilde{q}^2 F^{\prime \prime \prime \prime} (\tilde{q})] \Biggr) d \tilde{q}\quad .
\label{second I derivative}
\end{split}
\end{align}

To show that $I(q_{+})$ is a nondecreasing function of $q_{+}$ when conditions $1-5$ on $F$ hold we start in the linear limit where it is straight forward to show that $\partial_{q_{+}} I(q_{+}) = 0$. When conditions $1-5$ hold the integrand in equation (\ref{second I derivative}) is non-negative for all $q,q_{+}$. Thus $\partial^2_{q_{+}} I(q_{+}) \ge 0$ which implies that $\partial_{q_{+}} I(q_{+}) \ge 0$.

\section{Trigonometric Parametrisation for Numerical Calculations of Untwisted Modes} \label{trig reg}

As pointed out in \citet{Ogilvie19} the presence of an apparent singularity when $q = 0$ presents a problem for calculating untwisted modes numerically. These apparent singularities can be removed by using trigonometric parametrisation \citep{Ogilvie19}. 

Starting with the Hamiltonian for the nonlinear oscillator (equation \ref{q Hamiltonian}) and making a canonical transform to canonical variables $(\beta,\pi_{\beta})$ given by

\begin{equation}
q =: \sin 2 \beta \quad , \pi_{\beta} := 2 \cos 2 \beta \pi_q \quad ,
\end{equation}
with Hamiltonian

\begin{equation}
H = \frac{\pi_{\beta}^2}{4 k^2 \cos^2 2 \beta (F^{\prime \prime}(\sin 2 \beta))^2} + F(\sin 2 \beta) \quad .
\end{equation}

For $\gamma=1$

\begin{equation}
F(\sin 2 \beta) = -2 \ln \cos \beta \quad, \qquad F^{\prime \prime}(\sin 2 \beta) = \frac{1}{2} (\sec^2 \beta \sec 2 \beta + 2 \sec 2 \beta \tan \beta \tan 2 \beta) \sec 2 \beta
\end{equation}
and $\gamma = 2$ 2D disc

\begin{equation}
F(\sin 2 \beta) = \sec 2 \beta \quad, \qquad F^{\prime \prime}(\sin 2 \beta) = (2 - \cos 4 \beta) \sec^5 2 \beta \quad .
\end{equation}

For each $q_{+}$ (and associated $\beta_{+}$) the nonlinear oscillator describing the $\varphi$ dependence can be solved via a shooting method to find the value of $k$ which makes the solution periodic. This gives us $k$ as a function of $q_{+}$ from which we can obtain the integral $I(q_{+})$ from

\begin{equation}
I(q_{+}) = \frac{\pi}{\sqrt{2} k} \quad .
\end{equation} 

$J(q_{+})$ can be obtained from its relationship with $I(q_{+})$:

\begin{equation}
J(q_{+}) = \int_0^{q_{+}} \frac{F^{\prime} (q)}{2} I(q) \, d q \quad .
\end{equation}

Alternatively we can directly compute the integrals $I$ and $J$ using the regularised forms for $F$ and $F^{\prime \prime}$. This still possesses an apparent singularity at $q = q_{+}$ which must be dealt with separately. Typically it is sufficient to simply excise this apparent singularity by truncating the integral at a $q$ sufficiently close to $q_{+}$.


\bsp	
\label{lastpage}
\end{document}